\documentclass[a4paper,11pt]{article}
\pdfoutput=1

\usepackage{jcappub}

\usepackage[T1]{fontenc} 

\title{\boldmath First results of the cosmic ray NUCLEON experiment}

\author[a]{E. Atkin,}
\author[b]{V. Bulatov,}
\author[b]{V. Dorokhov,}
\author[c,d]{N. Gorbunov,}
\author[b]{S. Filippov,}
\author[c,d]{V. Grebenyuk,}
\author[e]{D. Karmanov,}
\author[e]{I. Kovalev,}
\author[e]{I. Kudryashov,}
\author[e]{A. Kurganov,}
\author[e]{M. Merkin,}
\author[e,1]{A. Panov,\note{Corresponding author.}}
\author[e]{D. Podorozhny,}
\author[b]{D. Polkov,}
\author[c]{S. Porokhovoy,}
\author[a]{V. Shumikhin,}
\author[e]{L. Sveshnikova,}
\author[c,f]{A. Tkachenko,}
\author[c,d]{L. Tkachev,}
\author[e]{A. Turundaevskiy,}
\author[e]{O. Vasiliev and}
\author[e]{A. Voronin}

\affiliation[a]{National Research Nuclear University ``MEPhI'',\\ Kashirskoe highway, 31. Moscow, 115409, Russia}
\affiliation[b]{SDB Automatika,\\ Mamin-Sibiryak str, 145, Ekaterinburg, 620075, Russia}
\affiliation[c]{Joint Institute for Nuclear Research,\\ Dubna, Joliot-Curie, 6, Moscow region, 141980, Russia}
\affiliation[d]{``DUBNA'' University,\\ Universitetskaya str., 19, Dubna, Moscow region, 141980, Russia}
\affiliation[e]{Skobeltsyn Institute of Nuclear Physics,\\ Moscow State University, 1(2), Leninskie gory, GSP-1, Moscow, 119991, Russia}
\affiliation[f]{Bogolyubov Institute for Theoretical Physics,\\ 14-b Metrolohichna str., Kiev, 03143, Ukraine}

\emailAdd{evatkin@mephi.ru}
\emailAdd{bulat@horizont.e-burg.ru}
\emailAdd{dvs-rtf@yandex.ru}
\emailAdd{gorbunov@sunse.jinr.ru}
\emailAdd{serg1812@mail.ru}
\emailAdd{greben@jinr.ru}
\emailAdd{karmanov68@mail.ru}
\emailAdd{im.kovalev@physics.msu.ru}
\emailAdd{ilya.kudryashov.85@gmail.com}
\emailAdd{me@sx107.ru}
\emailAdd{merkinm@silab.sinp.msu.ru}
\emailAdd{panov@dec1.sinp.msu.ru}
\emailAdd{dmp@eas.sinp.msu.ru}
\emailAdd{danilka.85@mail.ru}
\emailAdd{porokh@nusun.jinr.ru}
\emailAdd{shuma.v.v@mail.ru}
\emailAdd{tfl10@mail.ru}
\emailAdd{avt@jinr.ru}
\emailAdd{tkatchev@jinr.ru}
\emailAdd{ant@eas.sinp.msu.ru}
\emailAdd{oav@rsx.sinp.msu.ru}
\emailAdd{voronin@silab.sinp.msu.ru}

\keywords{cosmic ray experiments, particle acceleration, ultra high energy cosmic rays}

\abstract{The NUCLEON experiment was designed to study the chemical composition and energy spectra of galactic cosmic ray nuclei from protons to zinc at energies of $\sim10^{11}$--$10^{15}$\,eV per particle. 
The research was carried out with the NUCLEON scientific equipment installed on the Russian satellite ``Resource-P'' No.\,2 as an additional payload. 
This article presents the results for the measured nuclei spectra related to the first approximately 250 days of the scientific data collection during 2015 and 2016. 
The all-particle spectrum and the spectra of p, He, C, O, Ne, Mg, Si and Fe are presented. 
Some interesting ratios of the spectra are also presented and discussed. 
The experiment is now in its beginning stage and the data still have a preliminary character, but they already give numerous indications of the existence of various non-canonical phenomena in the physics of cosmic rays, which are expressed in the violation of a simple universal power law of the energy spectra. 
These features of the data are briefly discussed.}

\begin{document}
\maketitle
\flushbottom

\section{Introduction}
\label{sec:intro}

\begin{figure}
\centering
\includegraphics[width=0.75\textwidth]{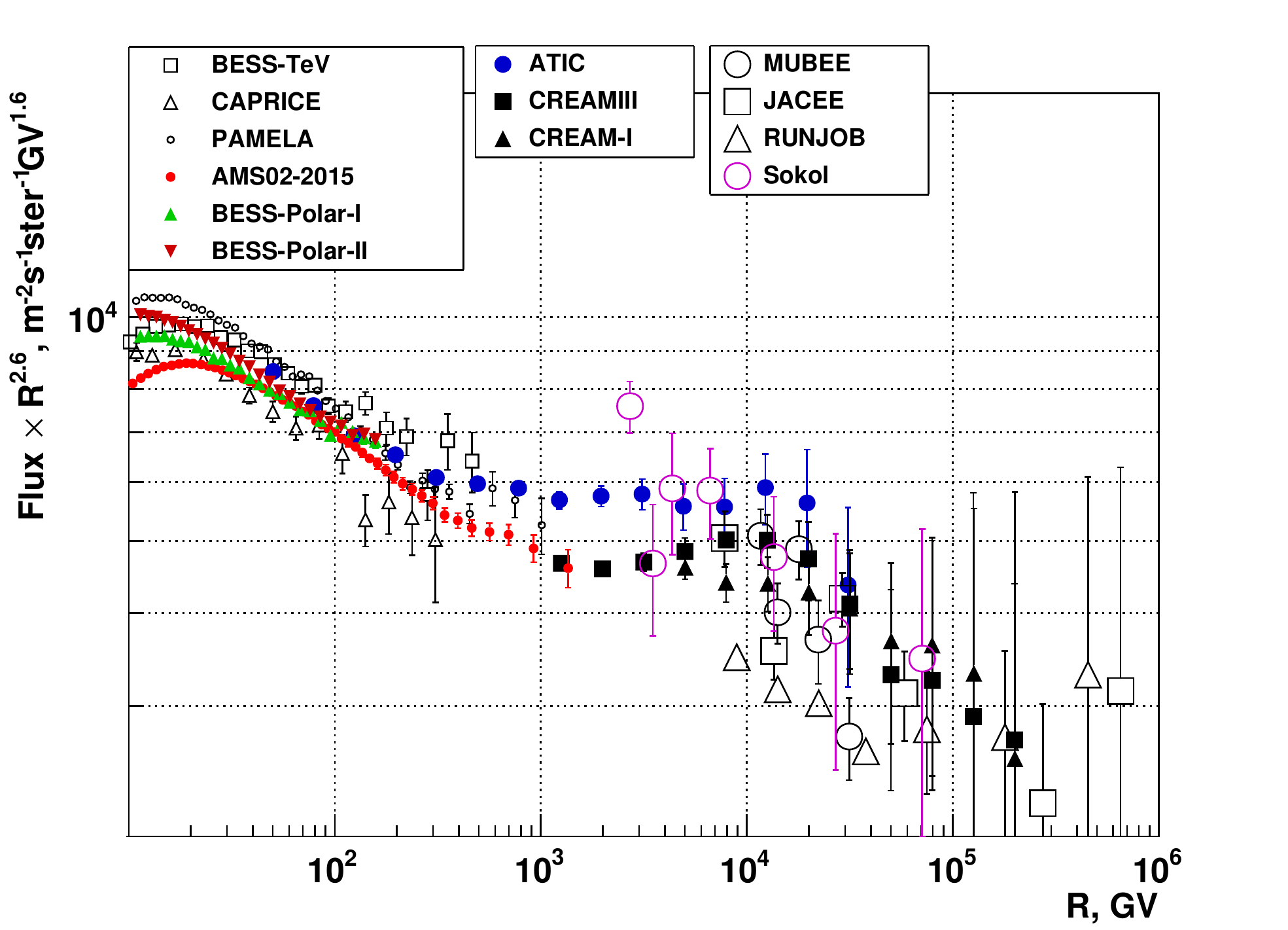}
\caption{\label{fig:Compilation-p} A compilation of data on the measurement of the proton spectrum of cosmic rays by direct experiments: BESS-TeV \cite{BESS-TeV-2003,BESS-TeV-2004,BESS-TeV-2005}; CAPRICE \cite{CAPRICE-2003}; PAMELA \cite{CR-PAMELA-2011-p-He-Mag}; AMS02-2015 \cite{AMS-02-2015-PRL1}; ATIC \cite{ATIC-2009-PANOV-IzvRAN-ENG}; CREAM-III \cite{CREAM2017-ApJ-pHe}; CREAM-I \cite{CREAM2011-ApJ-PHe-I}; MUBEE \cite{MUBEE-1993-JetpLett,MUBEE-1994-YadFiz}; JACEE \cite{JACEE-1998-ApJ}; RUNJOB \cite{RUNJOB-2005-ApJ}; SOKOL \cite{SOKOL-1993-IzvRAN}.}
\end{figure}

One of the most notable features in the energy spectrum of cosmic rays is the sharp increase in the slope of the energy spectrum near $3\times10^{15}$\,eV (3 PeV) per particle -- the so-called ``knee.'' 
The nature of this ``knee'' is still unclear, and represents one of the major mysteries of cosmic ray physics and space physics in general. 
The ``knee'' in the spectrum of cosmic rays has been found and is still observed in the EAS (extensive air showers) experiments, which provide data on the energy spectrum of cosmic rays at very high energies, but do not give reliable information about their chemical composition. 
At the same time, for understanding the physics near the ``knee,'' it would be very important to know the behavior of the individual components of the flux of cosmic rays near this area. 
Much more detailed information on the chemical composition of cosmic rays is provided by so-called direct experiments, in which the spectrometer is moved out of the atmosphere to a stratospheric balloon or a spacecraft, where space particles can be observed directly, using different types of spectrometers. 
Such experiments provide indications of complex behavior of the spectra of individual components of cosmic rays at energies 10\,TeV -- 1\,PeV, i.e. in the region adjacent to the knee from the low-energy side, but such data are severely lacked and do not have sufficiently high statistical reliability. 
For example, figure~\ref{fig:Compilation-p} shows a short compilation of data on the measurement of the proton spectrum of cosmic rays by direct experiments. 
Firstly, there is a noteworthy feature in the form of upturn of the spectrum near the energy of $\sim$500 GeV, the presence of which is well established in several experiments, although the details of the behavior remain to be studied. 
Secondly, there is an indication of a break in the energy spectrum near 10\,TeV, but so far no experiment has given statistically reliable data in this respect. 
The behavior of the spectrum at energies above 100\,TeV is completely unclear. 
There is an urgent need to improve the quality of results for energies from several TeV up to about 1000 TeV. 
There are a number of examples of other similar problems in the energy spectra of other nuclei, which are given below.

\begin{figure}
\centering
\includegraphics[width=0.75\textwidth]{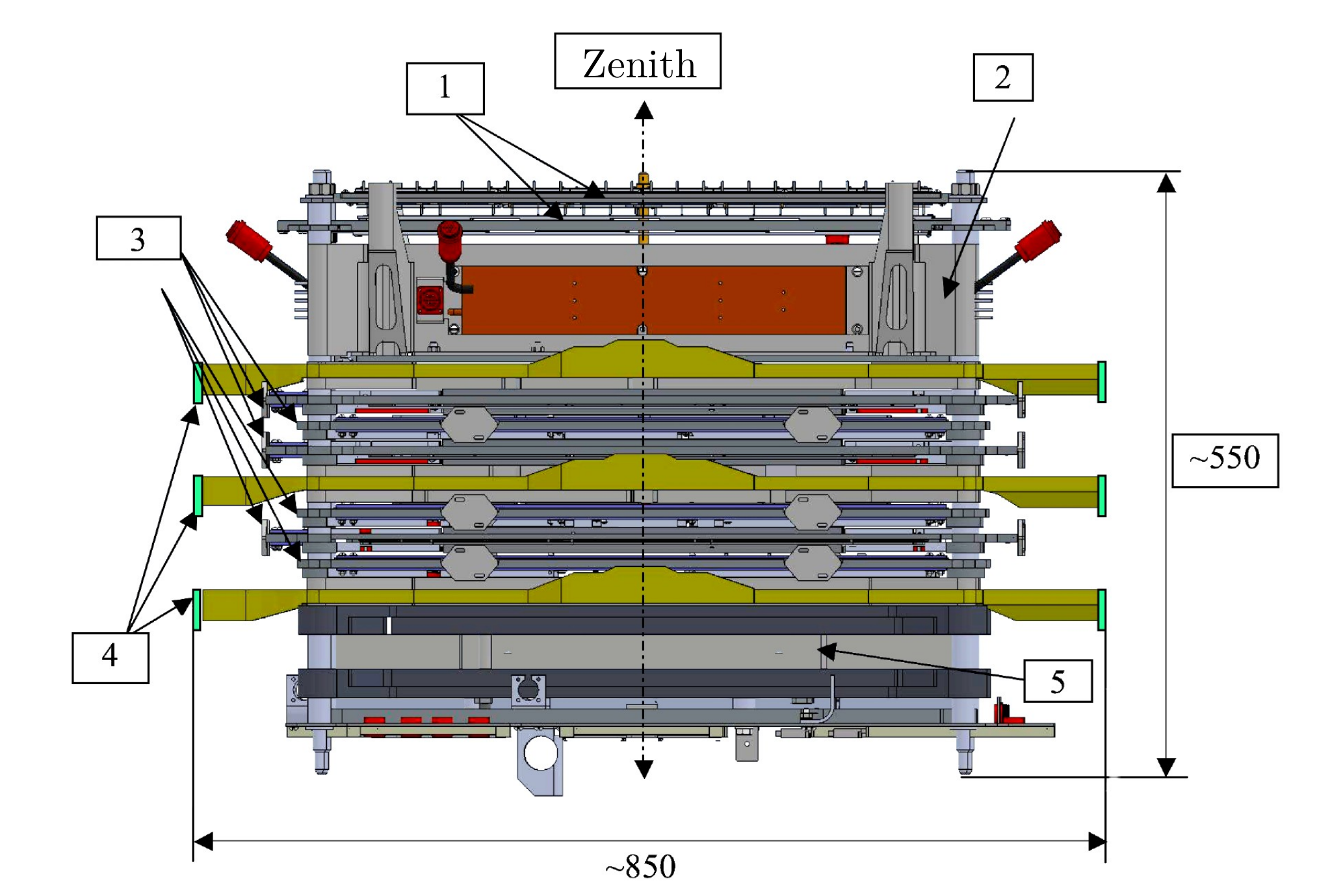}
\caption{\label{fig:NUCLEON} A simplified layout diagram of the NUCLEON spectrometer. 
1 -- two pairs of planes of the charge measurement system (ChMS); 2 -- a carbon target; 3 -- six planes of the energy measurement system using the KLEM method (KLEM system tracker); 4 -- three double-layer planes of the scintillator trigger system (the trigger system); 5 -- a small aperture calorimeter (IC).
}
\end{figure}

The NUCLEON experiment was designed primarily to solve the problems outlined above. 
Thus, the main priority of the NUCLEON experiment is to measure the spectra of cosmic ray nuclei with an individual charge resolution in the energy range from 10\,TeV to 1\,PeV per particle, while having a lower energy threshold of a few hundred GeV. 
This review presents the main results of the NUCLEON experiment concerning the energy spectra of cosmic ray nuclei obtained from a set of statistics in 2015--2016.

\section{Features of the NUCLEON detector}
\label{sec:features}

The NUCLEON experiment is a purely domestic project and has been developed with the participation of several institutions and universities in the Russian Federation. 
On December 28, 2014, the NUCLEON detector was launched into a sun-synchronous orbit with an average altitude of 475\,km and an inclination of 97~degrees as an additional payload of the Russian satellite Resource-P~2. 
On January 11, the NUCLEON detector was powered and started to collect data. 
The weight of the detector is approximately\,360 kg; the power consumption does not exceed 160\,W. 
The detector can transmit up to 10\,GB of scientific data to Earth per day. 
The planned lifetime of the NUCLEON detector is at least five years.

The most important feature of the NUCLEON detector is the implementation of two different particle energy measurement methods: the first uses an ionization calorimeter, and the second is a kinematic method, the Kinematic Lightweight Energy Meter (KLEM) \cite{KLEM-2000,KLEM-2001,KLEM-2002,KLEM-2005A,KLEM-2005B}, which is based on the measurement of the multiplicity of secondary particles after the first nuclear interaction of a primary particle with a target of the spectrometer. 
The first method is well known and has been used in experiments on cosmic rays many times. This is the first time the  second method has been used. 
The advantage of the use of two methods is the ability to cross-check the results of the measurements. 
The advantage of the KLEM method compared to the conventional calorimetric method is the ability to provide a high aperture of the device with a low weight of the equipment. 
The presence of the two methods of energy measurement in the NUCLEON detector will allow studying and calibrating the new KLEM method using a conventional calorimetric method.

Figure~\ref{fig:NUCLEON} shows a simplified diagram of the layout of the NUCLEON detector. 
The main systems of the spectrometer are two pairs of planes of the charge measurement system (ChMS), a carbon target, six planes of the energy measurement system using the KLEM method (KLEM system tracker), three double-layer planes of the scintillator trigger system, and a small aperture calorimeter (IC). 
Details of the detector design are provided in the articles \cite{NUCLEON-DEZ-2007A,NUCLEON-DEZ-2007B,NUCLEON-DEZ-2007C,NUCLEON-DEZ-2010,NUCLEON-DEZ-2015}.

\begin{figure}
\centering
\includegraphics[width=0.9\textwidth]{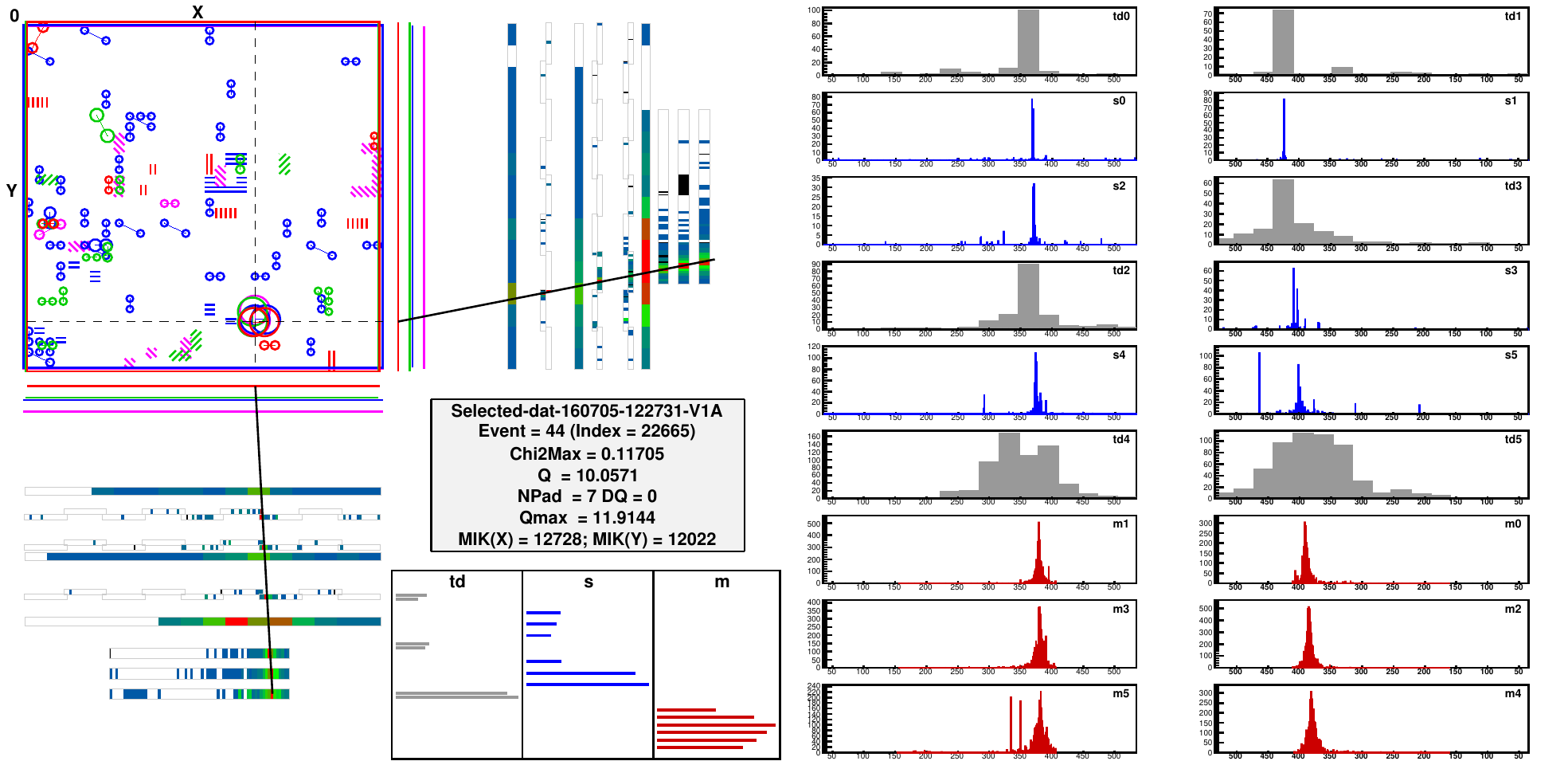}
\caption{\label{fig:Portrait} An example of an event visualization recorded by the detector (Ne nucleus).}
\end{figure}

Figure~\ref{fig:Portrait} shows an example of visualization of an event, recorded by the detector. 
The upper left corner shows a top view of the four planes of the ChMS, where each plane has its own color. 
The circles mark the signals from the triggered detectors, and the circle's area is proportional to the  effective charge measured by the detector. 
In addition to the primary particle signals, there are visible signals from the back scattered secondary particles as well as a certain amount of noise. 
Below and to the right of the ChMS planes the $XZ$ and $YZ$ (respectively) projections of the detector are displayed. 
The colors correspond to the value of the signal in the triggered detectors; black rectangles are inoperative detectors. 
The center of the figure shows a panel with some technical information about the event; below lie cascade curves obtained in the trigger system, the KLEM system and the IC (indicated in the figure as {\tt td}, {\tt s} and {\tt m}, respectively). 
A reconstructed shower axis is drawn over the projections. 
The right half of the figure displays the histograms of the energy released in planes of different systems.

\begin{figure}
\centering
\includegraphics[width=0.49\textwidth]{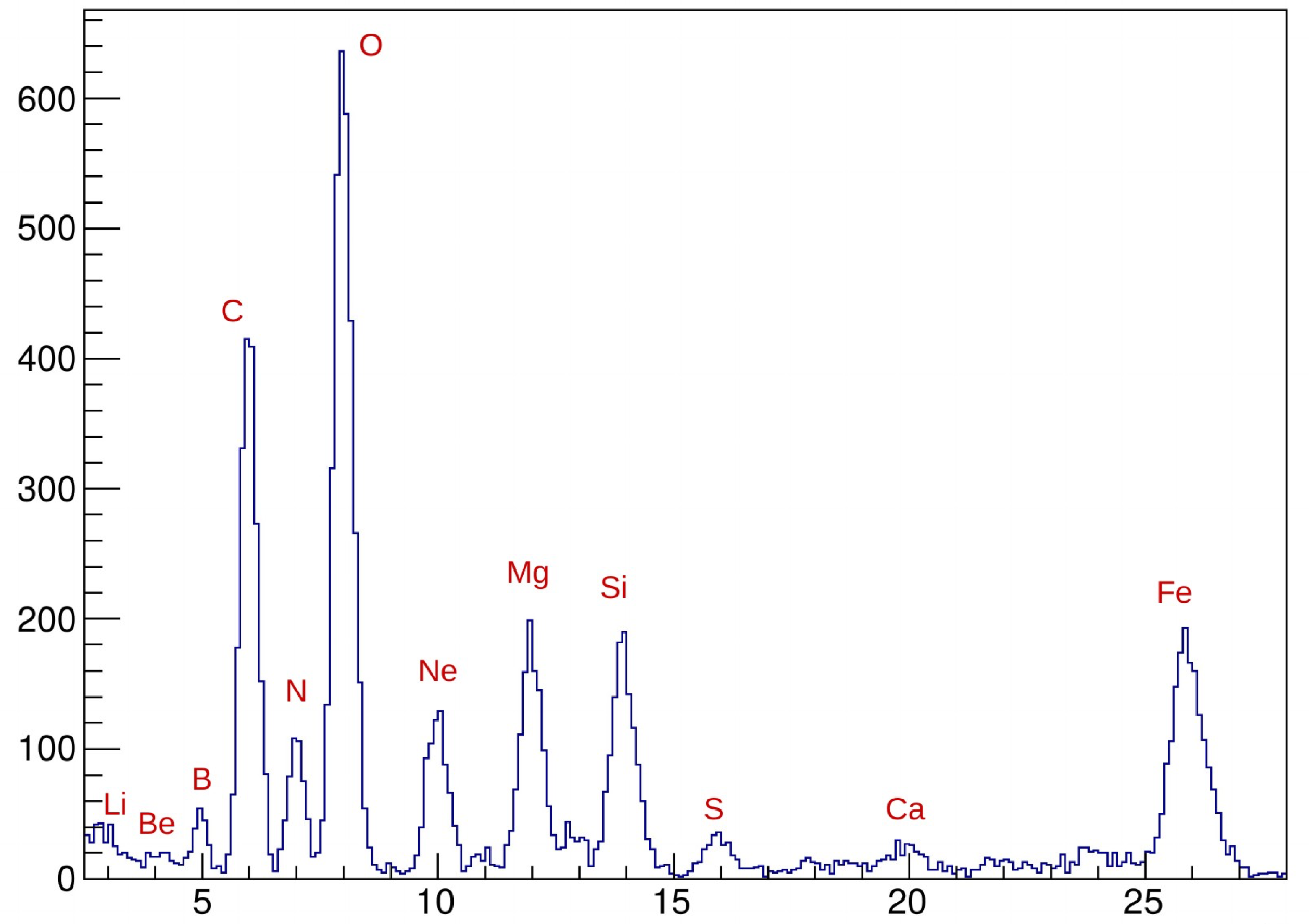}
\includegraphics[width=0.49\textwidth]{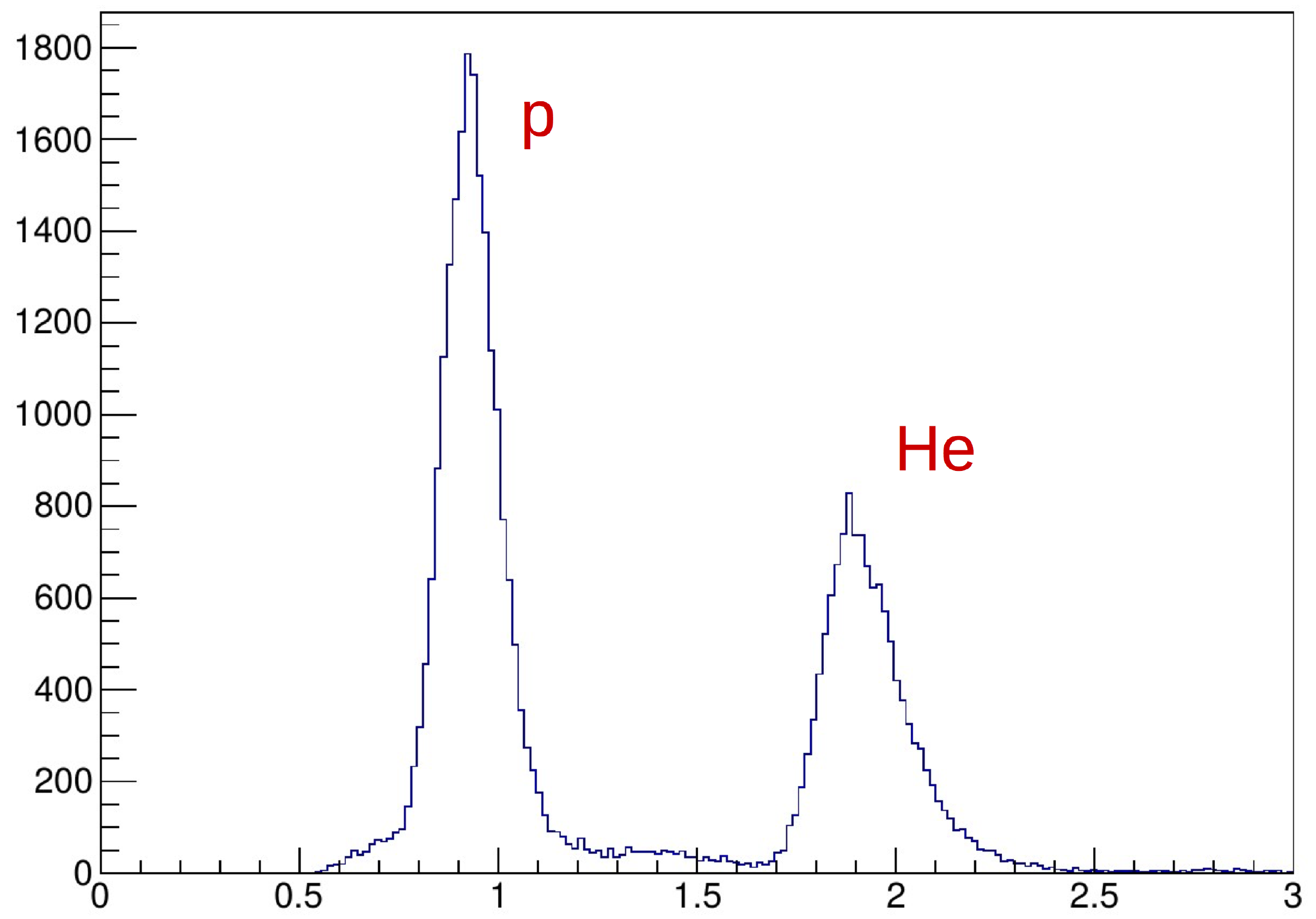}
\caption{\label{fig:Charge} Charge distributions of the cosmic ray nuclei measured in the NUCLEON experiment.}
\end{figure}

The ChMS system can reliably separate the charges of the abundant nuclei of cosmic rays to obtain their individual energy spectra. 
The charge distributions obtained by the NUCLEON detector are shown in figure~\ref{fig:Charge}.

Section \ref{sec:Results} presents the main results of the measurements of the cosmic ray energy spectra, for approximately one year of data-taking of the NUCLEON experiment. The results are shown for both energy measurement methods: the calorimetric method and the KLEM method. In each of these methods, there is a complex analyzing cycle before the final absolute energy spectra of cosmic rays is obtained. Some main steps of the implementation of both calorimetric and KLEM methods on board the NUCLEON spectrometer are discussed in Section~\ref{sec:EReconstruction}, but the methods in all detail will be published elsewhere in two special separate papers. 

The degree of consistency of the methods can be judged by the degree of consistency of the results. 
Here a very direct and model-independent argument is given in favor of expecting consistent results from both methods, that is, if all the data processing is performed correctly. 
The basic value which is used in the calorimetric method to reconstruct the spectra of the particles is the energy deposited in the detectors of the calorimeter $Ed$; and in the KLEM method, the main parameter is a specially constructed estimator $S$, which is related to the number of secondary particles with a high pseudorapidity after the first interaction (see Equation~(\ref{eq:S}) below, for details see also \cite{KLEM-2000,KLEM-2001,KLEM-2002,KLEM-2005A,KLEM-2005B}). 
\begin{figure}
\centering
\includegraphics[width=0.5\textwidth]{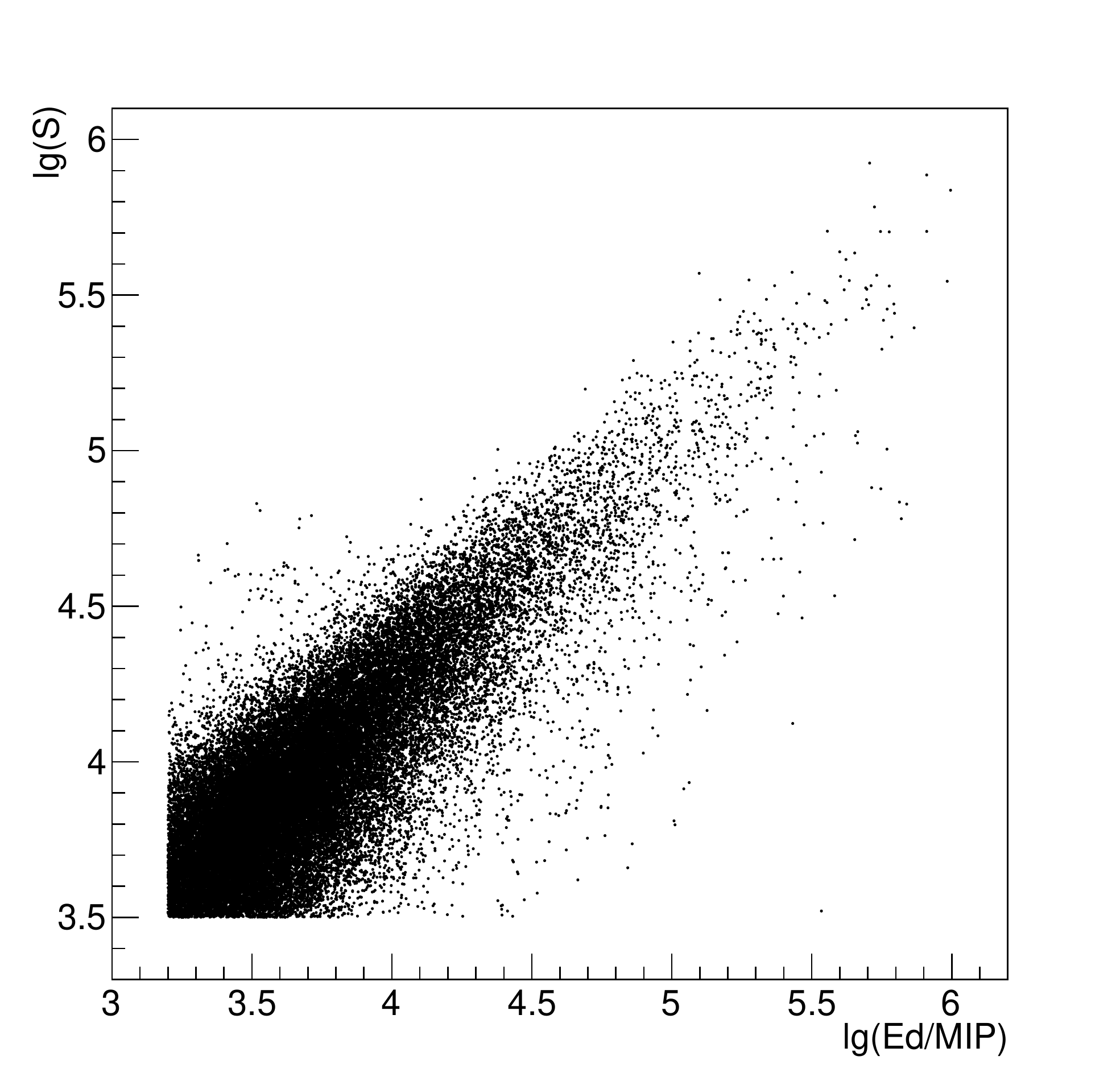}
\caption{\label{fig:LogEd-LogS} The scatter plot of the calorimeter energy deposit $Ed$ and the estimator $S$ of the KLEM method (for the incident He nuclei). 
The energy deposit $Ed$ is measured in MIP's -- the energy deposit of a minimally ionizing particle $(Z=1)$ for the silicon strip detectors of the calorimeter.}
\end{figure}
Figure~\ref{fig:LogEd-LogS} shows a scatter plot of the $Ed$ and $S$ variables, measured for the same event. 
A strong correlation between both parameters is visible. 
Obviously, if one of the values can be used for the reconstruction of the energy spectrum of the particles, then the other can be used for the same purpose as well. 
It is well-known that the energy deposited in the calorimeter $Ed$ can be used in this way, hence the estimator $S$ of the KLEM method can be used to reconstruct the spectra of cosmic rays particle energies too.

\section{Main steps of the energy reconstruction methods}
\label{sec:EReconstruction}

As it have been already mentioned above, two different particle energy measurement methods were implemented in the NUCLEON design: the KLEM method, that has been used in the astroparticle physics for the first time, and more usual method of ionization calorimeter. The KLEM method is considered to be a main method of the NUCLEON experiment since it provides greater statistics than the calorimetric method.

To determine the energy spectrum of primary particles, two fundamentally different approaches can be used. In the first approach, for nuclei of a certain type an apparatus function that gives the probabilities of obtaining different energy deposites $Ed$ of the calorimeter or different KLEM parameter $S$ for each primary particle energy is calculated by a simulation of the device. Then the experimental spectrum of $Ed$ or $S$ is constructed, and a complete inverse problem for the primary particles energy spectrum is solved for them. Such a problem, as it is known, belongs to the class of ill-posed inverse problems.

In the second approach, the energy of the primary particle is reconstructed for each event separately. For this, two functions must be defined. The first determines the factor that should be used to convert $Ed$ or $S$ into the primary energy of a particle. This coefficient (generally speaking) will depend on the $Ed$ or $S$ parameters themselves and can be determined computationally using the computer model of the apparatus. The second function gives the probability of a particle registration depending on the primary energy found (registration efficiency). When the total registration efficiency for a given event is found, the event must be added to the spectrum of registered particles with a weight equal to the inverse of the registration efficiency. This will take into account the missed particles.

Each of the two approaches mentioned above can be implemented in two versions. In the first version, the apparatus functions or the energy conversion factors and efficiency are determined depending on the direction of the shower axis (with some degree of the details of the direction description), in the second variant all these functions are determined by averaging over the entire working aperture of the spectrometer. The first option requires a much larger amount of simulation to build the apparatus functions, but it is somewhat more accurate than the second one.

In the versions of the methods described below, the second of these two approaches is realized: event-by-event method of energy reconstruction in its simplest form -- with averaging of the energy conversion factors and registration efficiency over the spectrometer aperture. We consider this approach as the first approximation for the data processing methods of the NUCLEON experiment.

\subsection{KLEM method}
\label{sec:KLEM}

In the KLEM method the primary energy is reconstructed by registration of spatial density of the secondary particles after the first hadronic interaction. Six planes of the KLEM energy measurement system (tracker) is located under the carbon target of 0.24 proton nuclear interaction length. It is supposed that the new secondary particles are generated by the first hadronic inelastic interaction in the carbon target. Then, additional secondary particles are produced in the thin tungsten converters of KLEM energy measurement system by electromagnetic and hadronic interactions. To reconstruct the primary energy of the incident particle the following $S$-estimator is used:
\begin{equation}
 S = \sum\limits_{i=1}^N n_i\eta_i^2,
 \label{eq:S}
\end{equation}
where summation are over N position-sensitive detectors of a tracker layer located after the converter; $\eta_i = -\ln(r_i/2H)$, where $r_i$  is the distance from the shower axis to $i$-th position-sensitive strip detector in a tracker layer ($r_i$ means $x_i$ or $y_i$ depending on the orientation of the strips of the tracker layer), $n_i$ is estimated number of charge-one particles crossing the detector, and $H$ is the distance from the interaction point in the target. For the real apparatus we apply $H$ determined as the distance from the middle of the carbon target to the tracker layer. Each layer of the tracker produces its own value of $S$, but the most reliable data are produced by two lowest tracker layers. The systematic uncertainty related to the uncertainty in the position of the first hadronic interaction in the target is small in comparison with the physical fluctuations. A direct simulation shows only negligible increasing of RMS deviation of reconstructed energy if one neglects by the differences between the true interaction point and the position of the middle of the carbon target. The above-mentioned multiplication of secondaries in the tungsten converters make energy dependence $S(E)$ of the estimator steeper than for simple multiplicity in the first interaction.

\begin{figure}
\centering
\includegraphics[width=\textwidth]{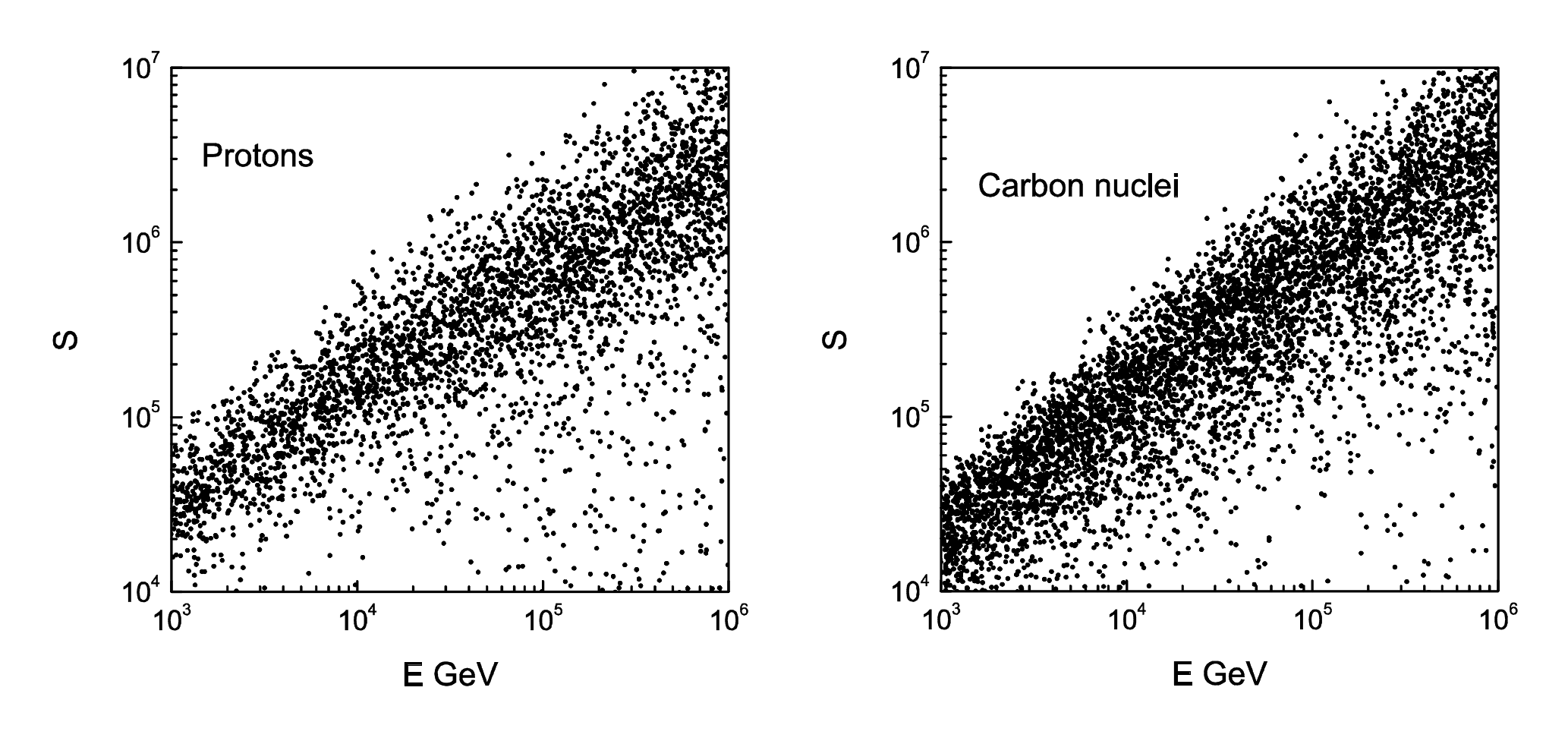}
\caption{\label{fig:KLEM-Scattering} The simulated scatter plots of the primary energy $E$ and the estimator $S$ for primary protons and carbon nuclei.}
\end{figure}

For an incident nucleus with mass number $A$ only a part of the nucleons interacts with the target carbon nucleus. Therefore, the multiplicity of secondaries is not proportional to $A$ but the angular distribution of secondaries is similar to the distribution for a proton. A detailed simulation of $S(E)$ for different nuclei is needed and have been performed by the GEANT 3.21 software package \cite{GEANT3-1984} complemented by the QGSJET \cite{KALMYKOV-1997} nuclear interaction generator to describe high-energy hadron-nucleus and nucleus-nucleus interactions. Generally, the $S(E)$ dependences for different types of primary nuclei is similar in the wide energy range and look like a simple power-law functions. Two examples of simulated scattering plots of the primary energy $E$ versus the estimator $S$ are shown in Figure~\ref{fig:KLEM-Scattering}.

To reconstruct the primary energy of a particle, the scatter plots like in Figure~\ref{fig:KLEM-Scattering} for different nuclei were approximated by power laws like
\begin{equation}
 E_{rec} = a(S\times10^{-5})^b
 \label{eq:KLEMFit}
\end{equation}
where the parameters $a$ and $b$ were optimized by the mean square method, proceeding from the requirement $\langle E_{rec}/E\rangle = 1$ for the given initial spectrum of projectile nuclei. The optimization procedure will be described in details elswhere. The values of $a,b$ for some nuclei, obtained for the initial power-law spectrum with the spectral index $\gamma = -2.6$, are shown in the Table~\ref{tab:KLEMab}.

\begin{table}
 \caption{\label{tab:KLEMab}The values of parameters $a,b$ for approximation Equation~(\ref{eq:KLEMFit}), obtained for the initial power-law spectrum with the spectral index $\gamma = -2.6$.}
\begin{center}
\begin{tabular}{|c|c|c|}
 \hline
 Projectile & $a$, GeV & b \\
 \hline
 p  & 1651  & 1.36 \\
 He & 2556  & 1.27 \\
 C  & 3514  & 1.18 \\
 S  & 4163  & 1.14 \\
 Fe & 4362  & 1.12 \\
 \hline
\end{tabular}
\end{center}
\end{table}

\begin{figure}
\centering
\includegraphics[width=0.49\textwidth]{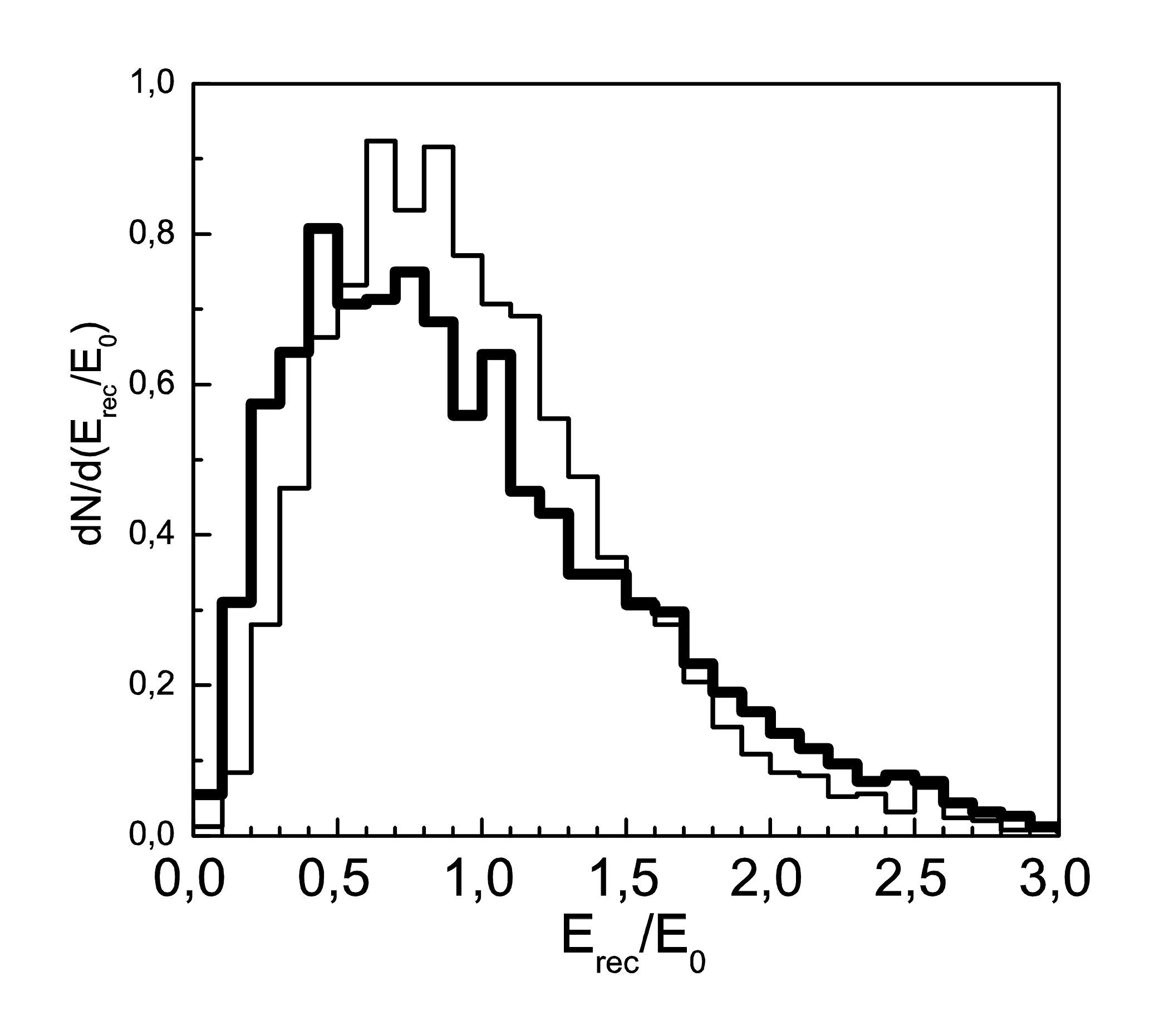}
\includegraphics[width=0.49\textwidth]{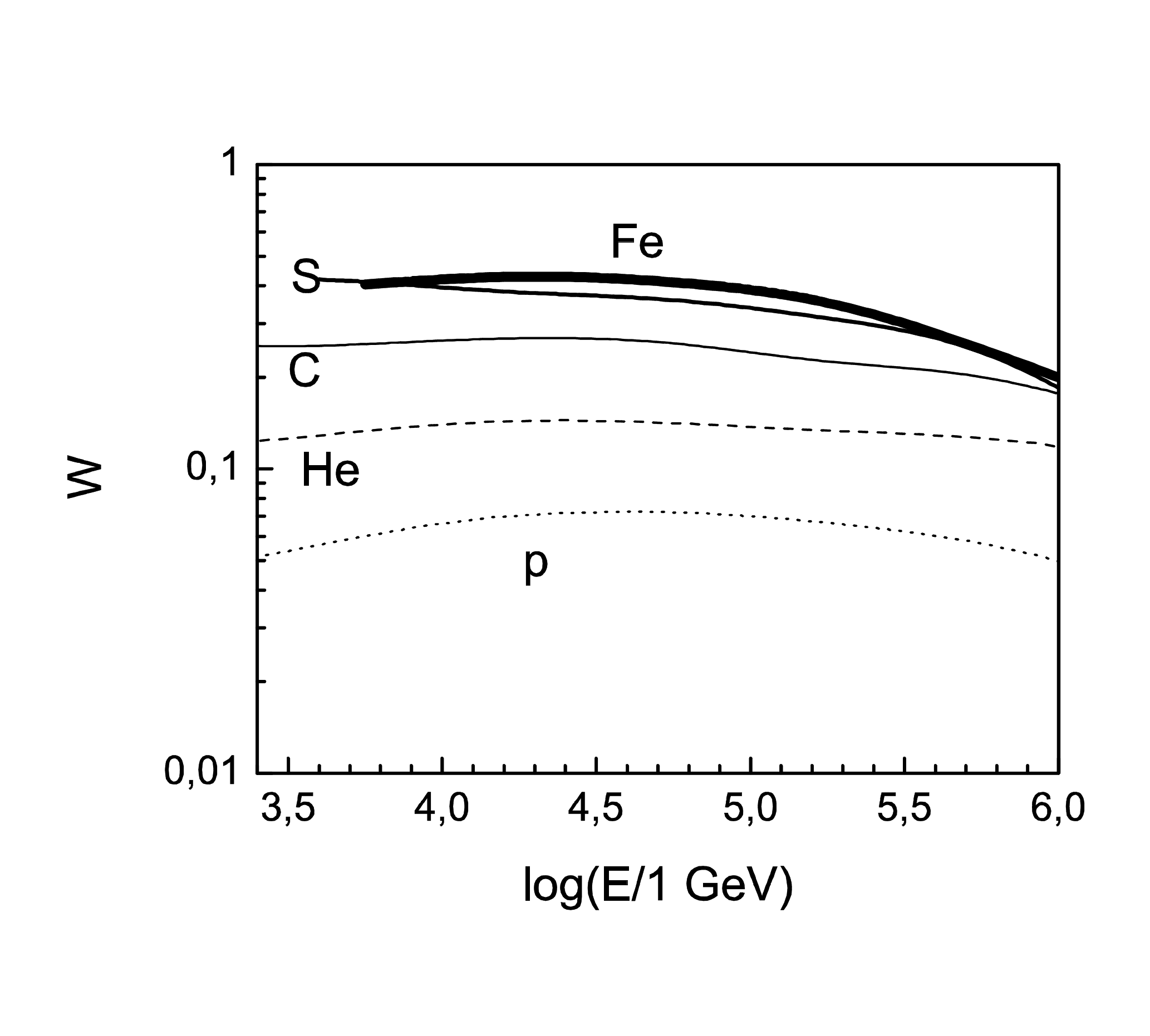}
\caption{\label{fig:KLEM-Reconstruct-Orig} Left panel: Normalized distributions of the reconstructed energy for primary pions with energies of 150 (thin line) and 350 GeV (thick line). Right panel: The energy dependences of the registration efficiency used in the KLEM method fore some nuclei.}
\end{figure}

The NUCLEON flight model was tested in 2012 on pion beams of the SPS accelerator in CERN. Pion data were obtained for 150~GeV and 350~GeV. The normalized distributions of the  energy, reconstructed by the KLEM method, for primary pions with energies of 150~GeV and 350~GeV are shown in Figure~\ref{fig:KLEM-Reconstruct-Orig}, the left panel. The RMS deviation to primary energy ratio is equal to 0.53 for 150~GeV and 0.63 for 350~GeV beams. The asymmetry of distributions is determined by the asymmetry of multiplicity distributions for hadron interactions. 

Within the framework of the present implementation of KLEM method, the determination of the efficiency of registration of particles as a function of the particle energy is considered as a separate problem. The energy dependences of the efficiency was calculated by simulation, according to the trigger conditions used. The calculated energy dependences of the registration efficiency for some nuclei and for one typical trigger condition are shown in Figure~\ref{fig:KLEM-Reconstruct-Orig}, the right panel. 

\subsection{Calorimetric method in the NUCLEON experiment}
\label{sec:MIC}

The idea of use of an ionization calorimeter for reconstruction of energy of cosmic-ray primary particles is based on the fact that the energy which is lost in a calorimeter by a shower is correlated with the energy of primary particle. Therefore the energy of a primary particle may be reconstructed with some accuracy from the energy, measured by the calorimeter.

Calorimeters can be divided on thick and thin. In thick calorimeters the shower caused by primary particle is absorbed almost completely. In such devices it is possible to reach a high precision of definition of energy of primary particles. In thin calorimeters the shower is absorbed not completely and the energy of a primary particle has to be determined only by a part of primary energy, which was absorbed by the calorimeter. The precision of energy measurement in thin calorimeters is lower due to fluctuations of the absorbed part of a shower.

In addition, calorimeters are divided into homogeneous and sampling ones. In homogeneous calorimeters the absorber is also an active medium that measures the deposited energy of the shower particles. In such devices, all the energy released in the calorimeter is measured. The sampling calorimeters contain a passive absorber in which, in fact, the nuclear and electromagnetic shower generated by the primary particle develops, as well as the detectors, which now measure not the energy, deposited in the calorimeter, but a value approximately proportional to the amount of ionizing particles in the shower. This value correlates with the energy deposit and, consequently, with the initial particle energy. Generally, homogeneous calorimeters provide higher accuracy.

The ionizing calorimeter IC of the NUCLEON spectrometer is a thin sampling calorimeter. The directly measurable quantity is the energy deposited in the thin silicon strip detectors (1\,mm  step) arranged in six layers between the layers of tungsten alloy (8\,mm thick each). The radiation depth of the calorimeter is 12 X-units, the nuclear depth of the calorimeter is 0.50 proton nuclear interaction lengths, the complete nuclear depth of the spectrometer from the top to the bottom is 1.12 proton interaction lengths.

\begin{figure}
\centering
\includegraphics[width=\textwidth]{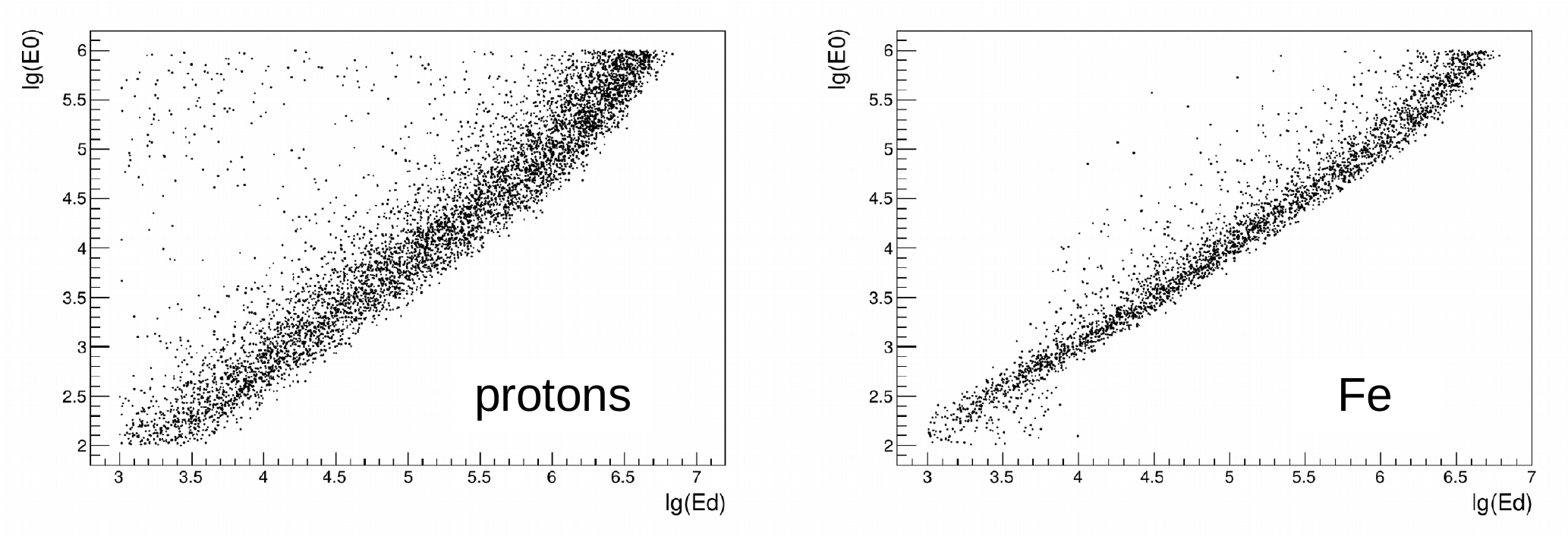}
\caption{\label{fig:MIK-ScatPlots} Scatter plots of the deposited energy $(Ed)$ versus the initial energy of the particle $(E0)$ for the primary protons and iron nuclei.}
\end{figure}

The energy deposit in the strip detectors of the IC calorimeter is measured in MIPs (MIP, mean energy loss rate close to the minimum for an one-charged ionizing particle). Since IC is a thin and, moreover, sampling calorimeter, the relationship between the energy of the primary particle and the energy measured by the calorimeter is of a statistical nature. The scatter plots of the deposited energy $(Ed)$ versus the initial energy of the particle $(E0)$ for the primary protons and iron nuclei obtained by simulation the NUCLEON spectrometer by the FLUKA system \cite{BATTISTONI-2015} are shown in Figure~\ref{fig:MIK-ScatPlots}. 
It is seen that the average correlation plots $E0$-$Ed$ does not lie on a simple power law. In particular, bending upward correlation curves is seen at the highest energy end of the plots. This phenomenon is associated with saturation of the electonics of the spectrometer detectors at the level above 27,000 MIPs per one strip detector, which sometimes arises at the highest primary particle energies. This saturation is taken into account in the simulation and is taken into account in the reconstruction of the energy of the primary particle.

The energy deposit $Ed$ of the calorimeter is recalculated to the initial energy of the particle $E0$ using a coefficient, that depends on $Ed$. $Ed$, expressed in MIPs, should be divided by this coefficient in order to obtain $E0$ in GeV -- this is the definition of this coefficient. The corresponding function, which is denoted as $K(Ed)$, was calculated for eight nuclei: p, He, Be, C, O, Mg, Ca, Fe, for which the interaction with the spectrometer was simulated, and for the remaining nuclei it was determined by interpolation in atomic weight.

Since the energy resolution of the IC calorimeter is not high, in order to calculate the most probable value of the conversion function $K(Ed)$ for each $Ed$, it is necessary to make an assumption about the shape of the initial cosmic-ray energy spectrum. It is known that the energy spectrum of all cosmic-ray nuclei in the energy region below $10^{15}$\,eV is close to the power-law fuction with an exponent of about $-2.6$ with some variations. It was this form of the spectrum that was supposed to be the initial approximation (this step is quite similar to that in the described above KLEM method). The calculation procedure for $K(Ed)$ is as follows. All the relevant area of the energy deposits $Ed$ is divided into relatively narrow bins. The initial flux of particles with the spectrum $\sim E^{-2.6}$ is simulated. For each $Ed$ bin, the distribution function for the ratios $Ed/E0$, which are the estimates of $K(Ed)$ for each individual event, is constructed. The most probable values of $K(Ed)$, which can be calculated from the histograms obtained, are used as the conversion factors from $Ed$ to $E0$.

\begin{figure}
\centering
\includegraphics[width=\textwidth]{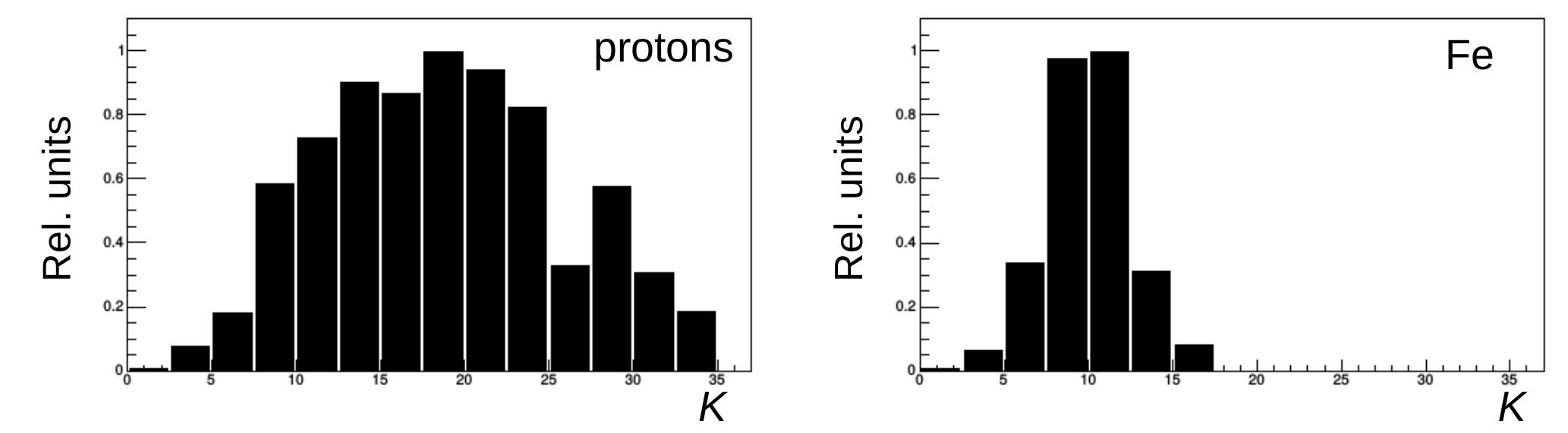}
\caption{\label{fig:MIK-KHist} $K(Ed)$-histograms obtained by simulation for the primary protons and iron nuclei for $Ed$-bin $5.0 < \lg(Ed/\mathrm{MIP}) < 5.5$. This energy bib corresponds to the primary energy of $\sim 10$\,TeV for protons and $\sim 18$\,TeV for iron nuclei.}
\end{figure}

In Figure~\ref{fig:MIK-KHist} two examples of $K(Ed)$-histograms obtained by simulation for the primary protons and iron nuclei are shown. The widths of the distributions obtained are good estimates of the energy resolution of the calorimetric technique in the NUCLEON experiment. The resolution is about 50\% for protons, and it is better for iron nuclei ($\sim$35\%).

\begin{figure}
\centering
\includegraphics[width=\textwidth]{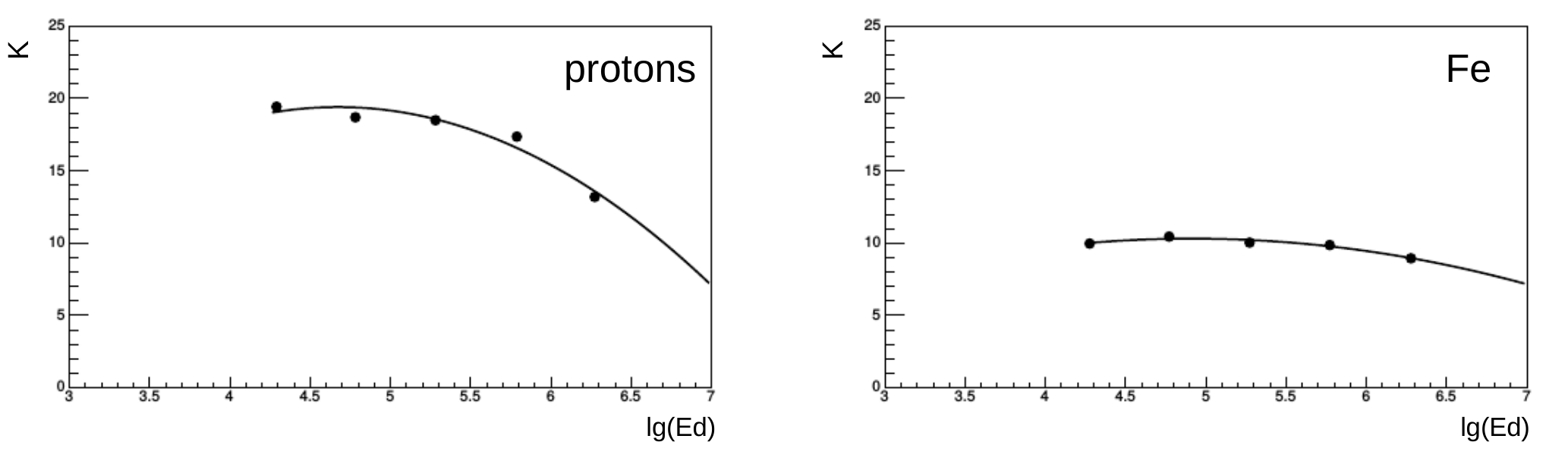}
\caption{\label{fig:MIK-KCalibr} $K(Ed)$ factors for protons and iron calculated for one of the most widely used flight trigger conditions.}
\end{figure}

According to the estimates of the most probable coefficients $K(Ed)$ for different $Ed$ and for each primary nucleus, quadratic interpolations of the corresponding functions are carried out. In Figure~\ref{fig:MIK-KCalibr} $K(Ed)$ factors for protons and iron calculated for one of the most widely used flight trigger conditions are given as an example. The curves are far from horizontal lines, which indicates that there is no proportionality between $Ed$ and $E0$, although there is certainly a strong correlation, in the form of a functional dependence.

The efficiency of registration is deterimined by the three main factors: the efficiency of the trigger, the efficiency of reconstruction of the shower axis, and the efficiency of determining the particle charge. In addition, there are a number of less significant factors that we will not discuss here. 

The inefficiency of the trigger is determined by the fact that the energy release in the planes of the trigger system does not always exceed the set of thresholds of the triggers, which are known from the calibration of the trigger system. This can happen either because the initial energy of the particle was not high enough, either because the first nuclear interaction occurred too low in the instrument (below the carbon target) or a nucleus passed through the entire device without any nuclear interaction at all. The first circumstance establishes the natural lower energy threshold of the device, and the second leads to pure losses of statistics, which can occur at any initial particle energies.

The inefficiency of reconstruction of the trajectory and the inefficiency of determining the charge of the primary particle are determined by certain software limitations imposed on the quality of the reconstruction of the trajectory of the primary particle and on the degree of correspondence of the values of the charge signals obtained over different planes of the charge measurement system.

\begin{figure}
\centering
\includegraphics[width=\textwidth]{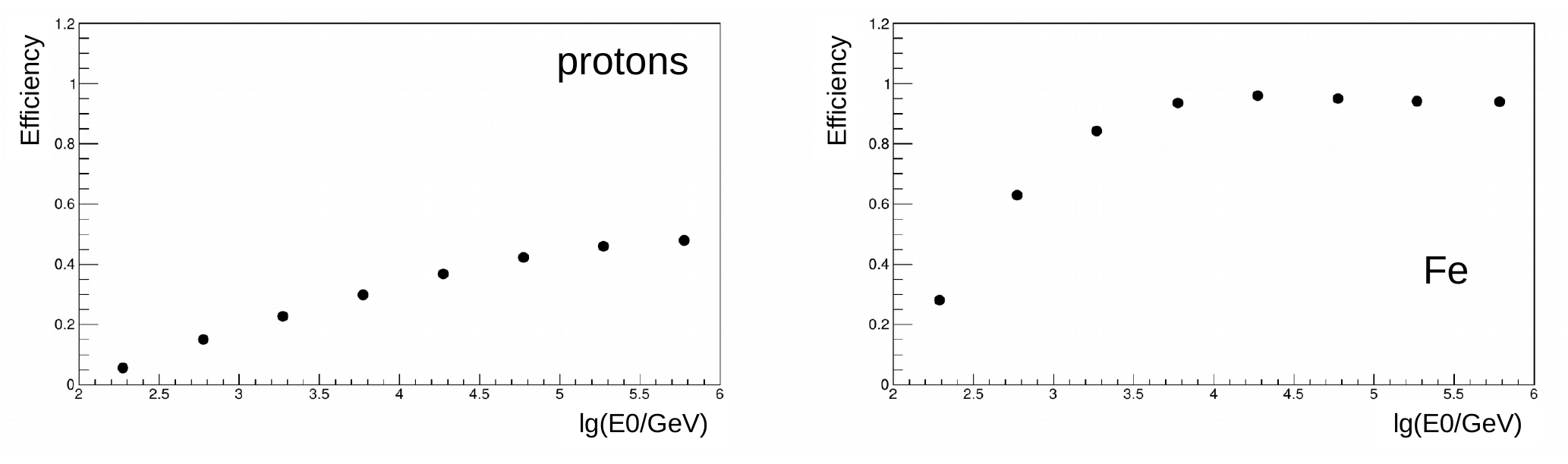}
\caption{\label{fig:MIK-Efficiency}. Efficiency curves for protons and for iron nuclei obtained by the simulation of spectrometer.}
\end{figure}

If Figure~\ref{fig:MIK-Efficiency} the efficiency curves for protons and iron obtained by the spectrometer simulation with accounting for all mentioned above factors are shown. It is seen that the efficiency for iron nuclei generally is higher than for protons, mainly due to larger nuclear cross-section of iron.

There are a number of less important factors (like accounting for a fraction of mistaken events etc.) in the reconstruction of the energy spectra both by the KLEM method and by the calorimetric method that we can not describe here in details due to a restricted volume of this paper. This issues will be described elsewhere in more special publications. In both the KLEM and the calorimetric methods the final energy spectrum of cosmic-ray nuclei is obtained as
\begin{equation}
 I(E) = \frac{N(E,\Delta E)}{T_l \times \Omega \times \Delta E \times R(E) \times Corr(E)},
\end{equation}
where $N(E,\Delta E)$ is the number of events near primary energy $E$ in the interval $\Delta E$, $T_l$ is the live time of the measurements, $R(E)$ is the efficiency of registration and $Corr(E)$ is a factor accounting for the mentioned above less important corrections.

\section{Main results}
\label{sec:Results}

This section will present the main results of the NUCLEON experiment spectra measurements for 2015--2016. 
Much of the time in this period was spent on the tests and the configuration of the detector, and part of the time was spent on a variety of technical manipulations of the Resource-P 2 spacecraft, during which data collection was not possible. 
The data presented correspond to 247 days of observations in terms of astronomical time, of which 160 days were the live time of the detector (the dead time was spent on the exchange of data between the detector and the on-board computer for event recording). 
The collected statistics are about one-fifth of the expected statistics, so the experiment is currently in its initial stage. 
The techniques for data processing at this stage of the experiment are also in the stage of checking, debugging, and partly even under construction, and therefore are preliminary. 
This is reflected in the nature of the reported results, which are also to be understood as preliminary. 
In particular, we do not try to give statistically accurate quantitative analyses of the data in this experimental phase, and, generally, we omit any detailed discussion of the physics of the observed phenomena (for a general discussion of the physics see section~\ref{sec:Discussion}). 
In the current phase of the research, that would be premature. 
The NUCLEON experimental data are, as a rule, given for two different energy measurement methods: the calorimetric and the KLEM methods. 
When comparing the results of the methods one should keep in mind that because its aperture is about four times larger, the KLEM method corresponds higher statistics than the calorimetric method.

\subsection{All-particle spectrum and the mean logarithm of atomic weight}
\label{subsec:AllPart}

\begin{figure}
\centering
\includegraphics[width=\textwidth]{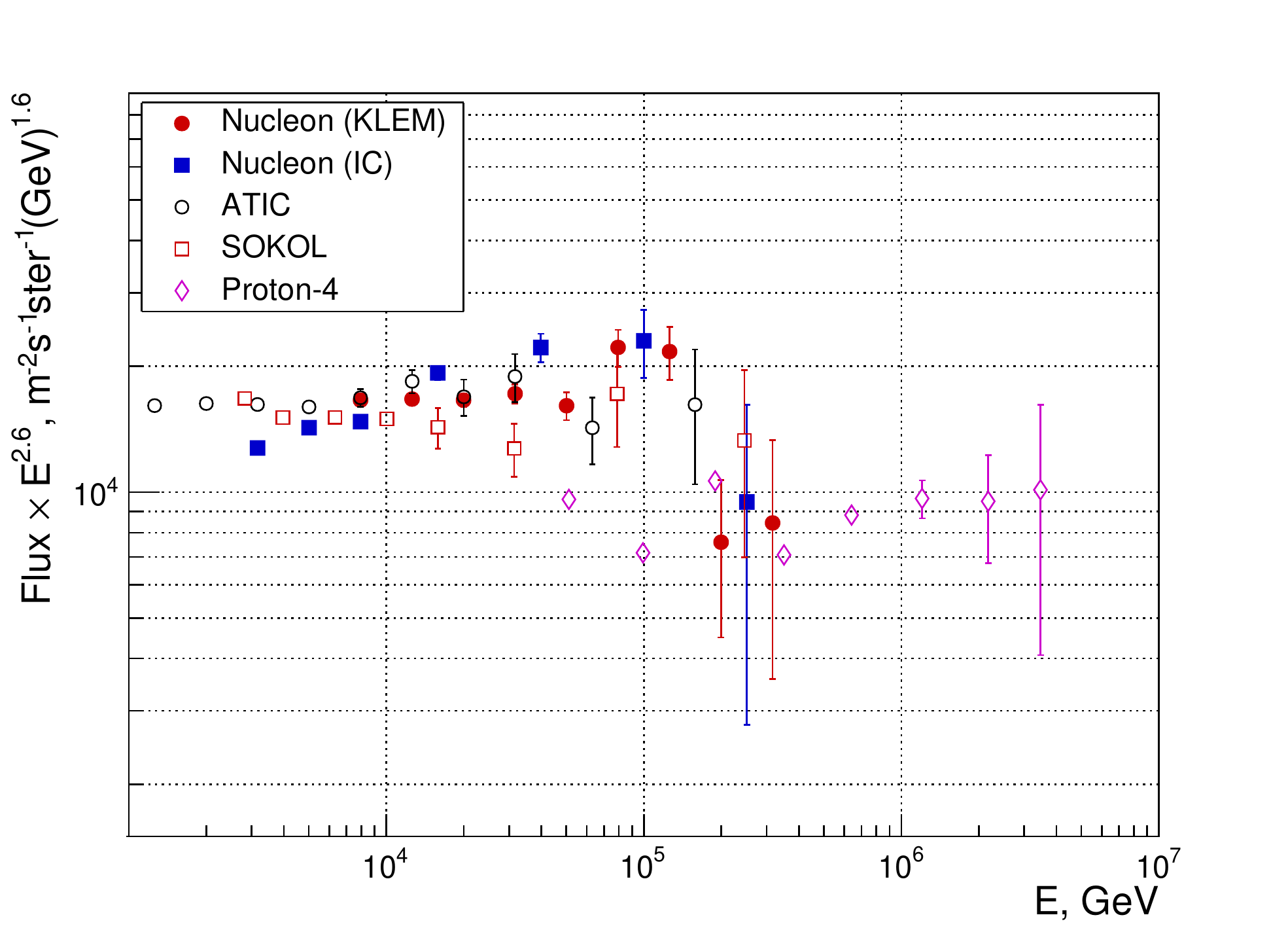}
\caption{\label{fig:All-Particles} All-particle spectrum measured by the KLEM system and by the calorimeter in comparison with the results of other direct measurement experiments: ATIC \cite{ATIC-2009-PANOV-IzvRAN-ENG}; Sokol \cite{SOKOL-1993-ICRC}; Proton-4 \cite{PROTON4-1972}.}
\end{figure}

\begin{figure}
\centering
\includegraphics[width=\textwidth]{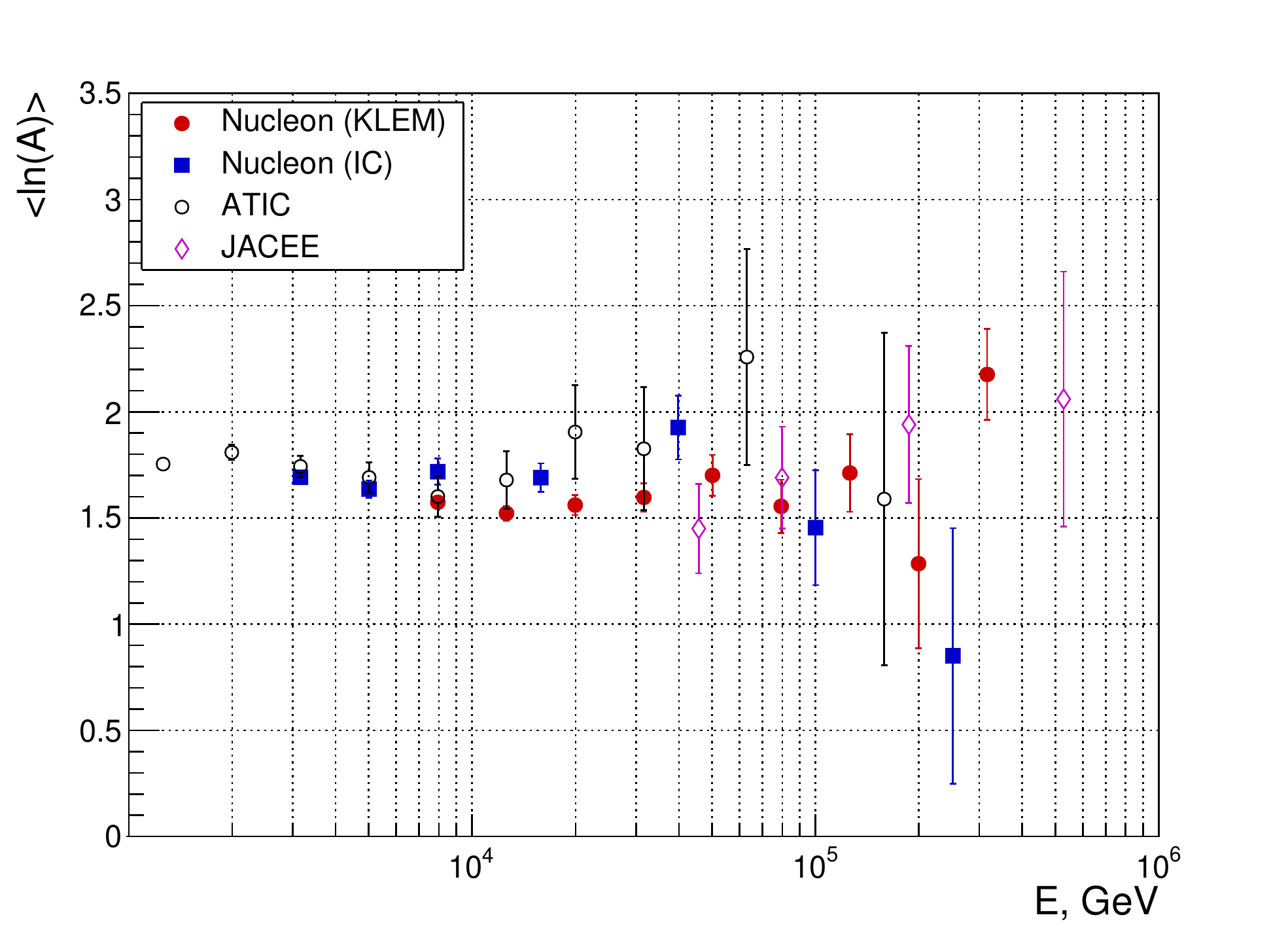}
\caption{\label{fig:LnA} A plot of the mean logarithm mass of cosmic rays versus energy per particle by the NUCLEON detector in comparison with the results of other direct measurement experiments: ATIC \cite{ATIC-2009-PANOV-IzvRAN-ENG}; JACEE \cite{JACEE-1998-NuclPhys}.}
\end{figure}

Figure~\ref{fig:All-Particles} shows the all-particle spectrum measured by the KLEM system and by the calorimeter in comparison with the results of other direct measurement experiments: ATIC \cite{ATIC-2009-PANOV-IzvRAN-ENG}, Sokol \cite{SOKOL-1993-ICRC}, and Proton-4 \cite{PROTON4-1972}. 
The spectrum for the KLEM method has a higher threshold than the spectrum of the calorimeter, as the KLEM system has not yet solved the problem of taking into account the so-called slips of the heavy nuclei. 
The problem is that a heavy nucleus, especially iron, may cause the trigger to activate a record of an event even without a nuclear interaction, by the ionization signals alone, as they are proportional to the charge squared, and therefore large for heavy nuclei. 
Such slips simulate an event with an initial energy of several TeV, and therefore it is necessary to work with a threshold higher than this energy. 
This problem can be solved, but in the current version of the data processing algorithms, it has not been solved yet. 
Since heavy nuclei in the KLEM are measured with a high threshold, the lower limit of the range of the all-particle spectrum can only be built up to highest value of the threshold of the individual nuclei.

The NUCLEON experimental spectra are in reasonable agreement with the ATIC and the SOKOL experiments, but all the spectra are notably higher in intensity than the spectrum of the Proton-4 experiment. 
The Proton-4 experiment still holds the record of highest energy achieved in a direct measurement of the energy spectrum of cosmic rays, but it was one of the first space experiments carried out, with a very simplified procedure, from a modern point of view, and it might have a low accuracy.

At energies above 100 TeV, both methods, the calorimetric and the KLEM, indicate a possible break in the spectrum of all particles. 
However, the statistics in this region of the spectrum are not enough even for preliminary conclusions.

A discrepancy between the results of the KLEM method and the calorimeter method outside the statistical error in the NUCLEON experiment should be noted. 
This suggests that some systematic errors in the measurement of the spectra still occur, although they are not very large. 
This was expected, since at this stage of the NUCLEON experiment, many experimental methods are preliminary, and the results will be refined. 
No detailed evaluation of systematic errors has been performed, as it is premature. 
This observation is relevant to almost all of the results to be presented in this paper.

Figure~\ref{fig:LnA} shows a plot of the mean of the logarithms of the masses of the cosmic rays versus the energy per particle by the NUCLEON detector, which exactly corresponds to the all particle spectrum in figure~\ref{fig:All-Particles}. 
The ATIC experiment indicates, with low statistical significance, the existence of an undulating structure (bending) near the energy 10\,TeV per particle. 
The curves of the mean logarithm mass of the NUCLEON experiment do not contradict the existence of such a structure and also give some indication of its existence. 
As the data set grows, the statistical significance of the NUCLEON experiment findings will grow, and the existence of the structure will be confirmed or refuted.

\subsection{Proton and helium spectra}
\label{subsec:p-He}

\begin{figure}
\centering
\includegraphics[width=\textwidth]{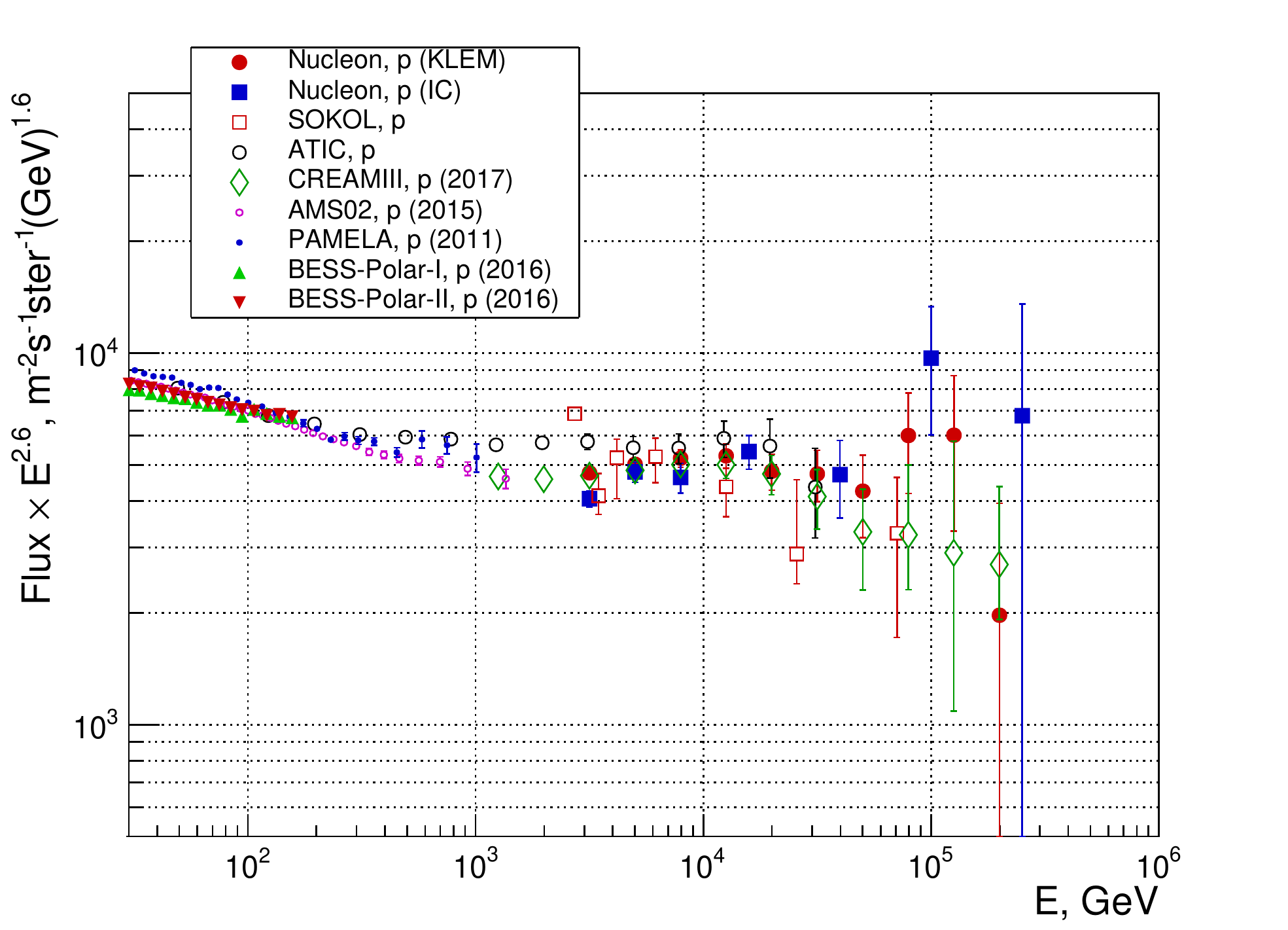}
\caption{\label{fig:p}Proton spectrum measured in the NUCLEON experiment together with the data from other experiments: Sokol \cite{SOKOL-1993-IzvRAN,SOKOL-1993-ICRC}, ATIC \cite{ATIC-2009-PANOV-IzvRAN-ENG}; CREAM-III \cite{CREAM2017-ApJ-pHe}; AMS-02 \cite{AMS-02-2015-PRL1};  PAMELA \cite{CR-PAMELA-2011-p-He-Mag}; BESS-Polar I and II \cite{BESS-Polar-2016}.}
\end{figure}

Figure~\ref{fig:p} shows a proton spectrum measured in the NUCLEON experiment together with the data from the Sokol \cite{SOKOL-1993-IzvRAN,SOKOL-1993-ICRC}, ATIC \cite{ATIC-2009-PANOV-IzvRAN-ENG}, CREAM-III \cite{CREAM2017-ApJ-pHe}, AMS-02 \cite{AMS-02-2015-PRL1},  PAMELA \cite{CR-PAMELA-2011-p-He-Mag} and BESS-Polar I and II \cite{BESS-Polar-2016} experiments. 
The results of the calorimetric and KLEM methods are close to each other and are in reasonable agreement with the results of the other experiments. 
However, it should be noted that there are discrepancies with the data of other experiments that are outside of the margins of statistical error, therefore they are methodological in nature. 
The proton spectra measured by the NUCLEON experiment do not contradict the existence of a break in the energy spectrum near 10 TeV, which was mentioned in the Introduction. 
Signs of the break with different statistical significance can be seen in the spectra of both the calorimeter and the KLEM. 
The behavior of the spectrum with energies above 100 TeV is unclear, as the statistics are insufficient, but these two methods do not exclude the spectrum's steepening after the break being replaced by a new flattening of the spectrum. 
The situation will become clearer with the collection of a larger set of statistics.

\begin{figure}
\centering
\includegraphics[width=\textwidth]{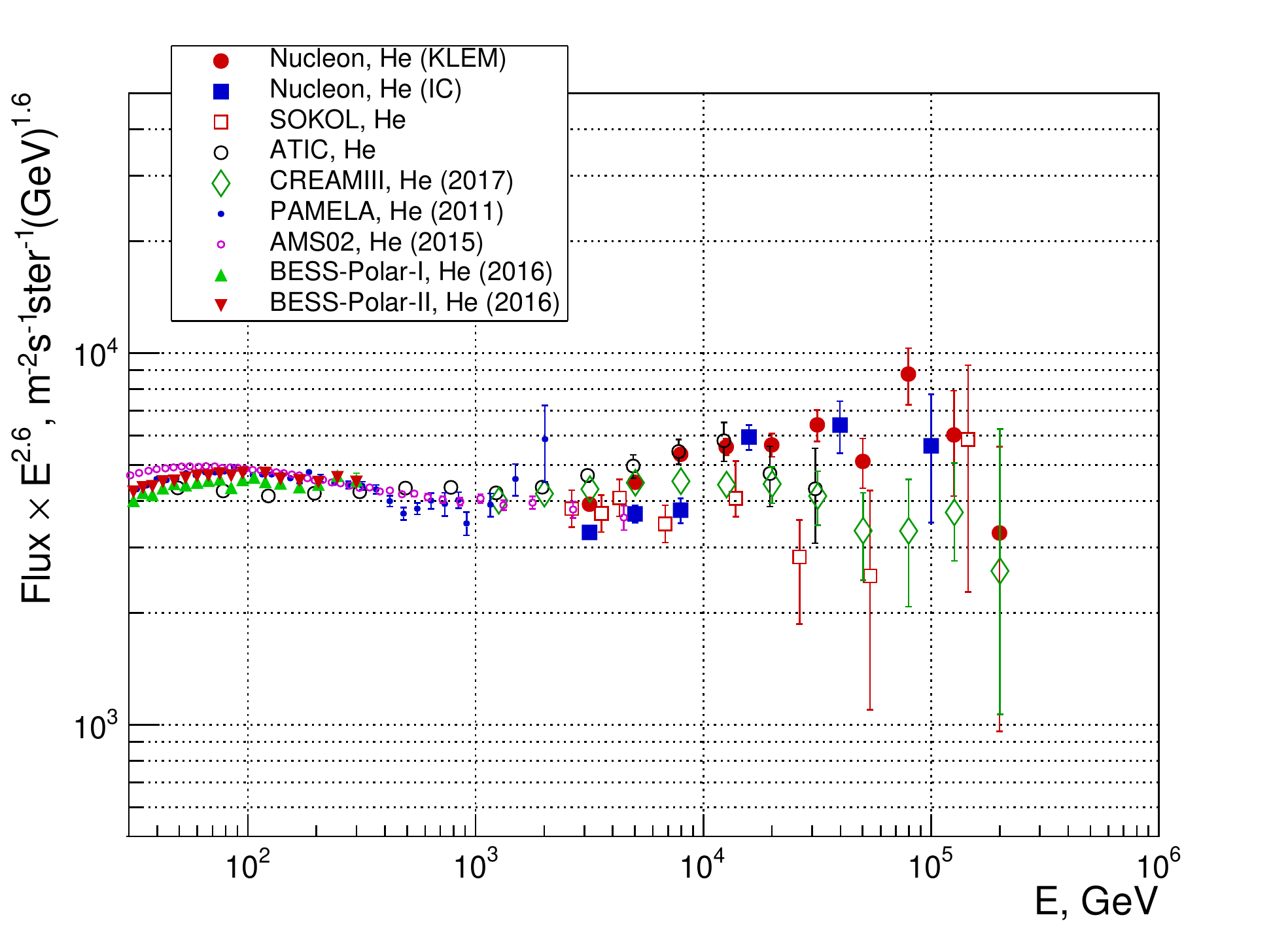}
\caption{\label{fig:He} Helium spectrum measured in the NUCLEON experiment together with the data from other experiments: Sokol \cite{SOKOL-1993-IzvRAN,SOKOL-1993-ICRC}, ATIC \cite{ATIC-2009-PANOV-IzvRAN-ENG}; CREAM-III \cite{CREAM2017-ApJ-pHe}; AMS-02 \cite{AMS-02-2015-PRL1};  PAMELA \cite{CR-PAMELA-2011-p-He-Mag}; BESS-Polar I and II \cite{BESS-Polar-2016}..}
\end{figure}

Figure~\ref{fig:He} shows the helium nuclei spectrum measured by the NUCLEON experiment, and the results of the Sokol \cite{SOKOL-1993-IzvRAN,SOKOL-1993-ICRC}, ATIC \cite{ATIC-2009-PANOV-IzvRAN-ENG}, CREAM-III \cite{CREAM2017-ApJ-pHe}, AMS-02 \cite{AMS-02-2015-PRL1},  PAMELA \cite{CR-PAMELA-2011-p-He-Mag} and BESS-Polar I and II \cite{BESS-Polar-2016} experiments. 
The NUCLEON data are consistent with other experiments. 
Some discrepancy may be noted for the two points of the Sokol experiment in the 20--50\,TeV range, but the statistical errors of the Sokol experiment are large, so this deviation is hardly a serious problem. 
At energies below 10\,TeV, a slight systematic difference between the calorimeter and the KLEM methods can be noted.

\begin{figure}
\centering
\includegraphics[width=0.6666\textwidth]{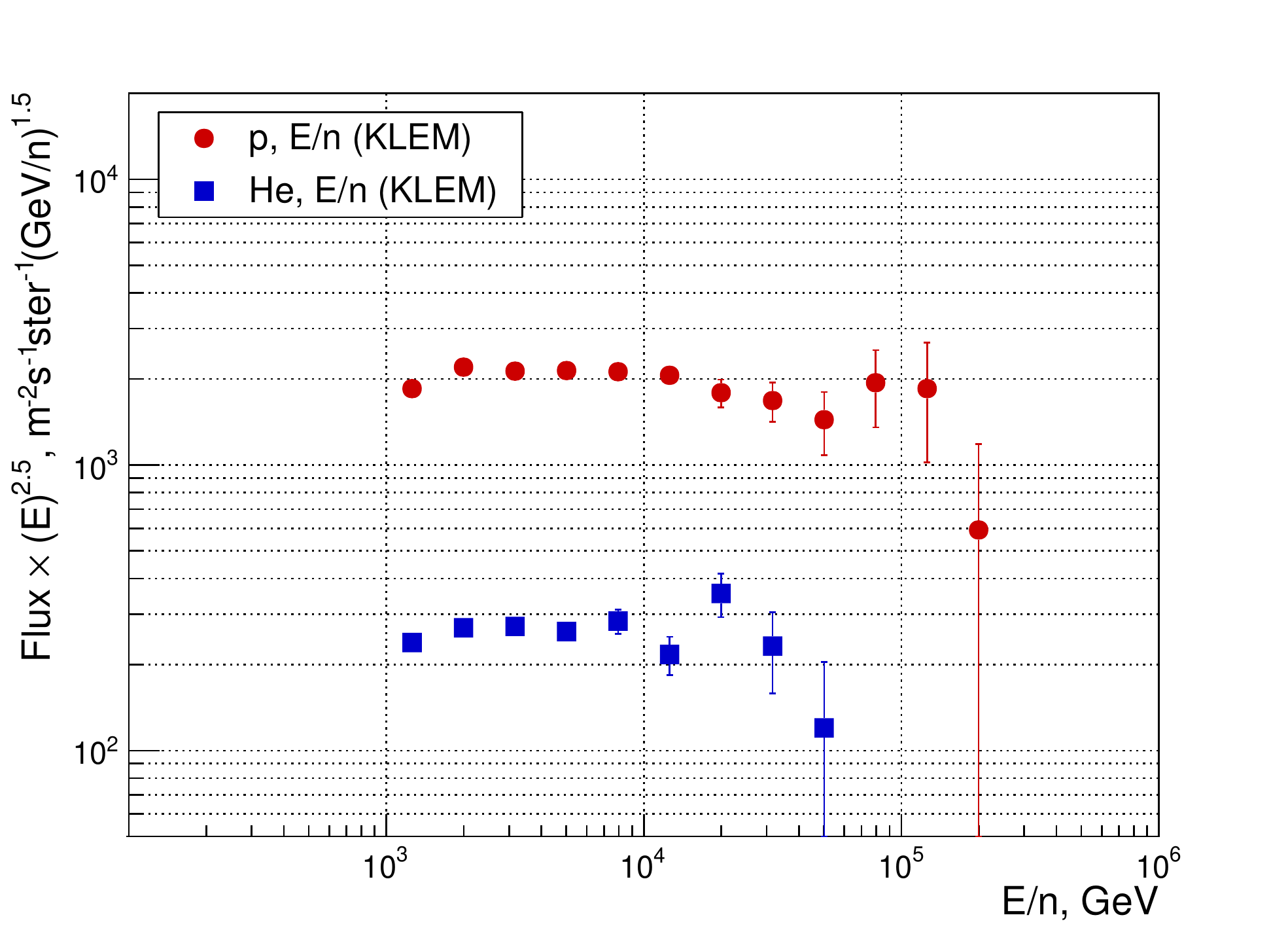}
\caption{\label{fig:p-He-KLEM}Spectra of protons and helium nuclei, measured in the NUCLEON experiment using the KLEM method in terms of energy per nucleon.}
\end{figure}

\begin{figure}
\centering
\includegraphics[width=\textwidth]{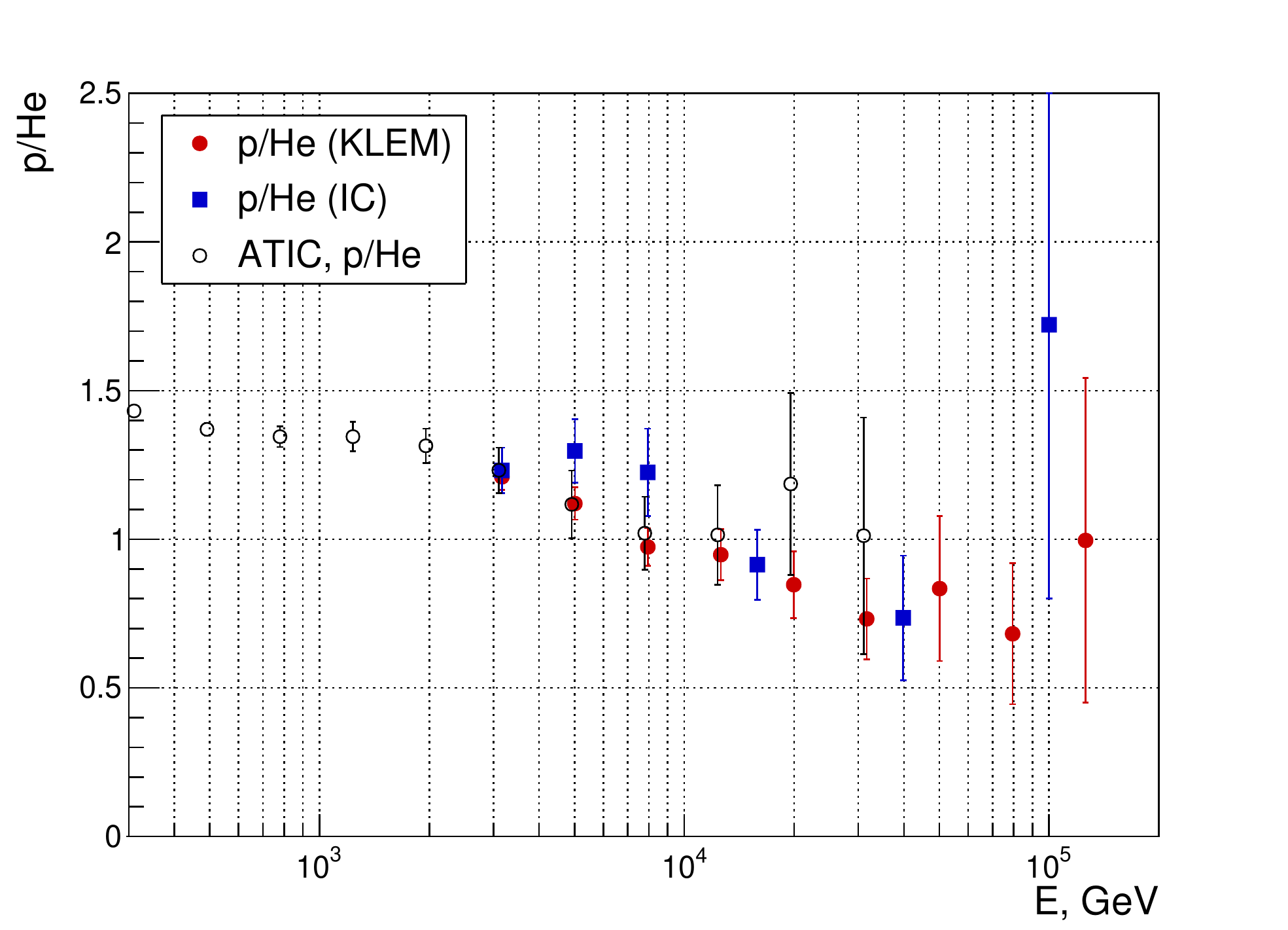}
\caption{\label{fig:p-to-He} Ratio of proton to helium flow of the NUCLEON experiment's KLEM and calorimeter methods and the data of the ATIC experiment \cite{ATIC-2009-PANOV-IzvRAN-ENG}.}
\end{figure}

Figure~\ref{fig:p-He-KLEM} shows the spectra of protons and helium nuclei measured in the NUCLEON experiment using the KLEM method in terms of energy per nucleon. 
The calorimetric method is not given here because it is qualitatively very similar to the KLEM results, but the statistics are worse. 
In figure~\ref{fig:p-He-KLEM}, a break is clearly visible in the proton spectrum near the energy of 10 TeV. 
In the comparison of the spectra of protons and helium in terms of energy per nucleon, it is noteworthy that the spectrum of helium gives some indication of a possible break at approximately the same energy as the break in the proton spectrum. 
This is a very interesting fact that should be carefully examined as the set of the NUCLEON experiment statistics grows.

Many direct experiments of the previous century gave an indication that the proton and helium spectra at energies ranging from tens of GeV to tens of TeV have different inclinations. 
This phenomenon would be of fundamental importance, as it would indicate different conditions of acceleration of protons and helium, and therefore the existence of different types of accelerators of cosmic rays. 
However, for a long time no experiment could give a statistically significant result in relation to the existence of such a difference, until the existence of the phenomenon with very high statistical reliability was confirmed in the region of energies from 200 GeV to 10 TeV in the ATIC experiment \cite{ATIC-2004-ZATSEPIN-IzvRan}. 
After that, the existence of the phenomenon was confirmed in several other experiments for various energy ranges, and for new experiments became in fact a test of a method's correctness. 
Figure 11 shows the ratio of proton to helium flux of the NUCLEON experiment's KLEM and calorimeter methods and the data of the ATIC experiment \cite{ATIC-2009-PANOV-IzvRAN-ENG}. 
The NUCLEON experiment confirms the presence of the phenomenon and its results are in full accordance with the results of the ATIC.

\subsection{Spectra of abundant heavy nuclei}
\label{subsec:Abundant}

\begin{figure}
\centering
\includegraphics[width=0.75\textwidth]{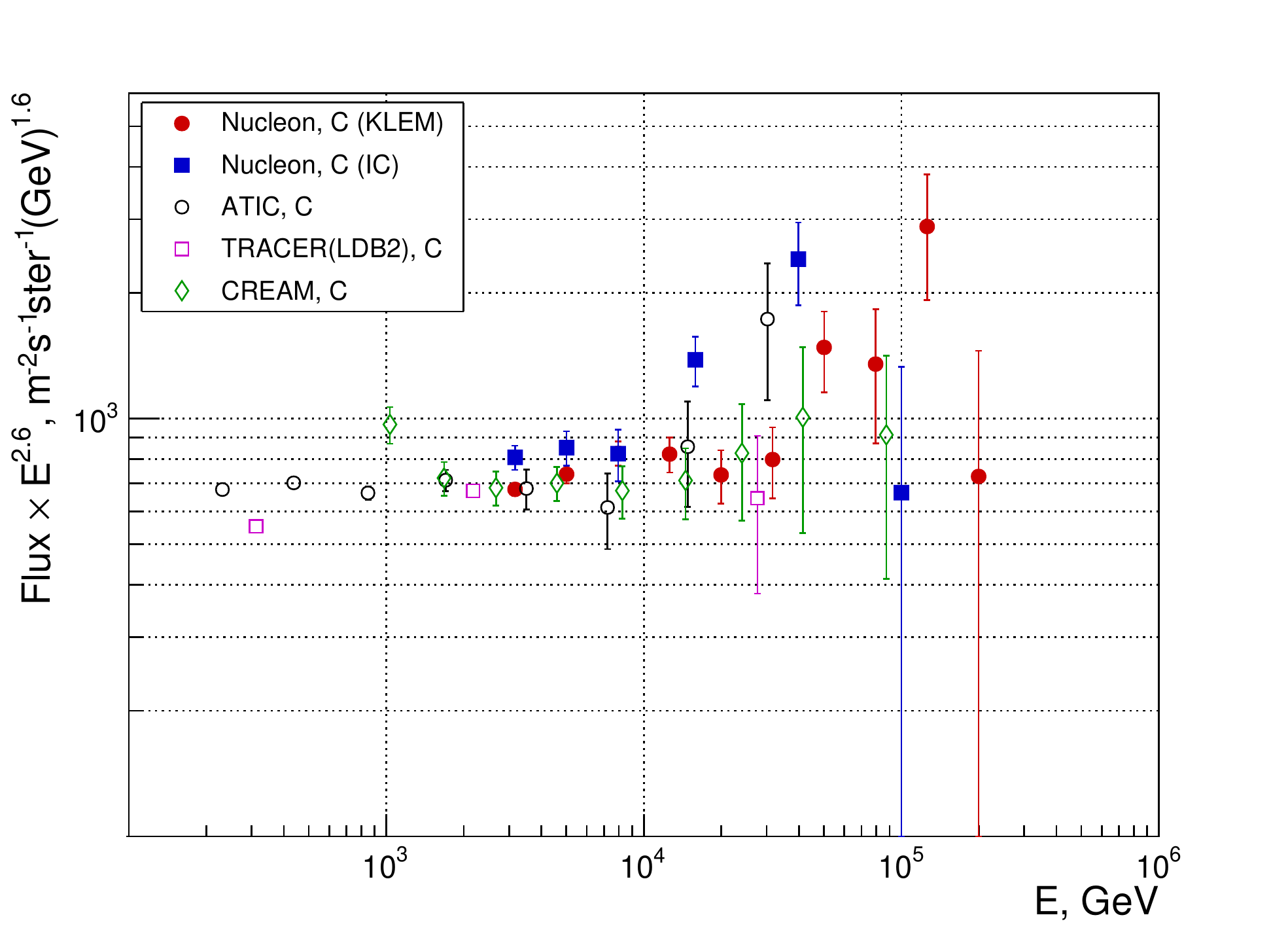}\\
\includegraphics[width=0.75\textwidth]{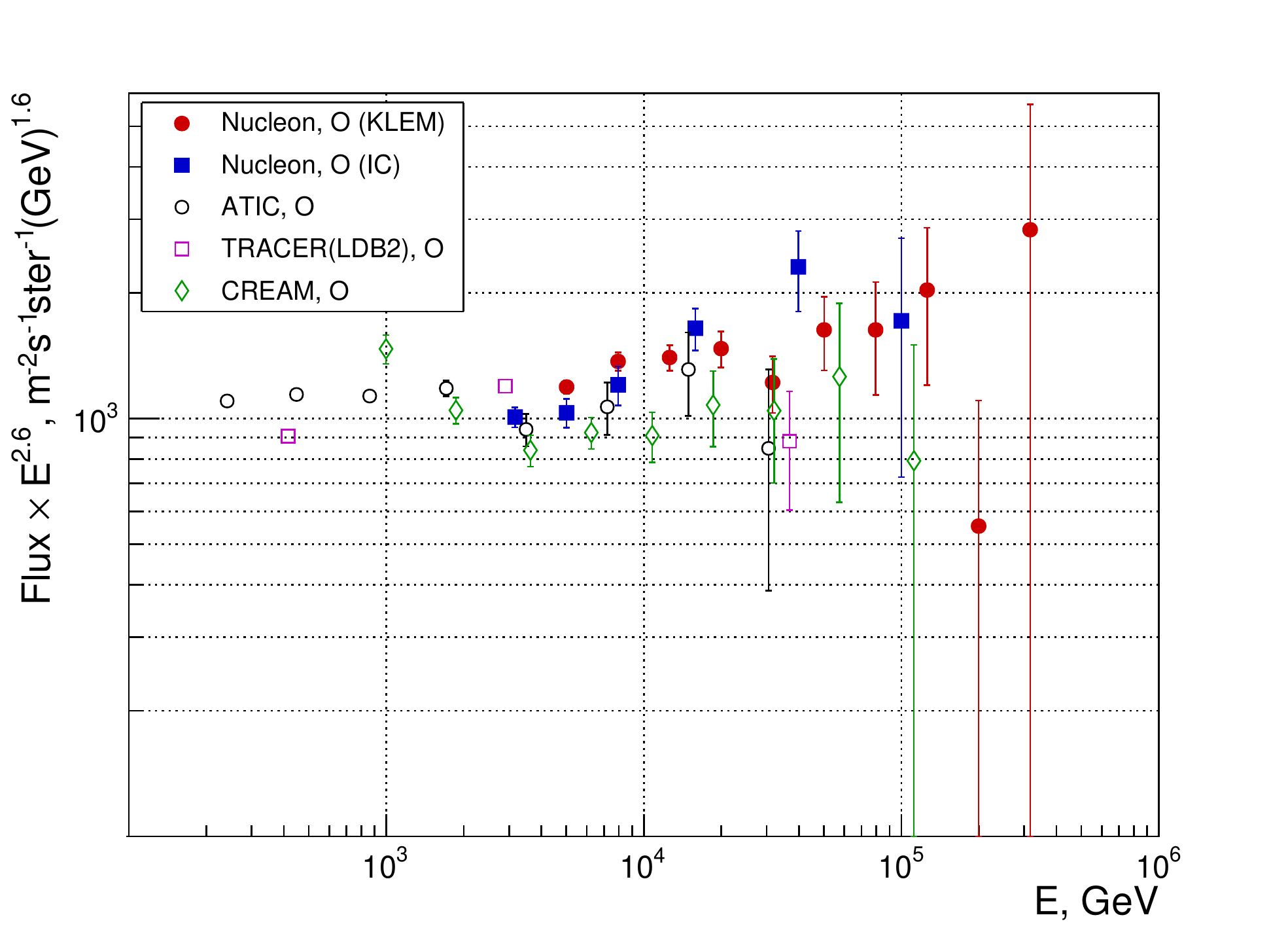}
\caption{\label{fig:CO} Energy spectra of carbon and oxygen nuclei obtained by the NUCLEON experiment and in the experiments ATIC \cite{ATIC-2009-PANOV-IzvRAN-ENG}, TRACER(LDB2) \cite{CRNUCL-TRACER2011-ApJ}, and CREAM \cite{CR-CREAM2010A}.}
\end{figure}

\begin{figure}
\centering
\includegraphics[width=0.75\textwidth]{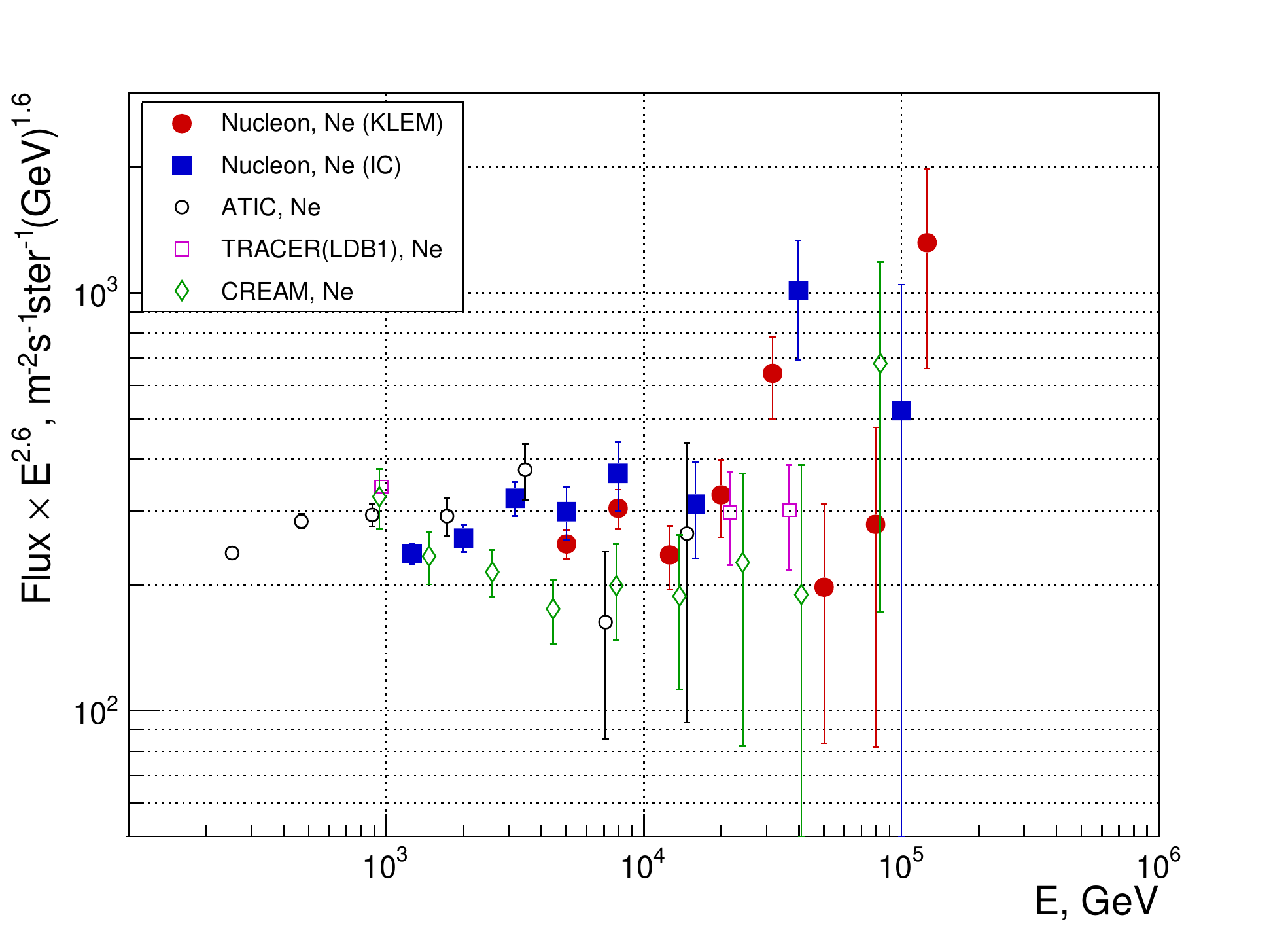}
\includegraphics[width=0.75\textwidth]{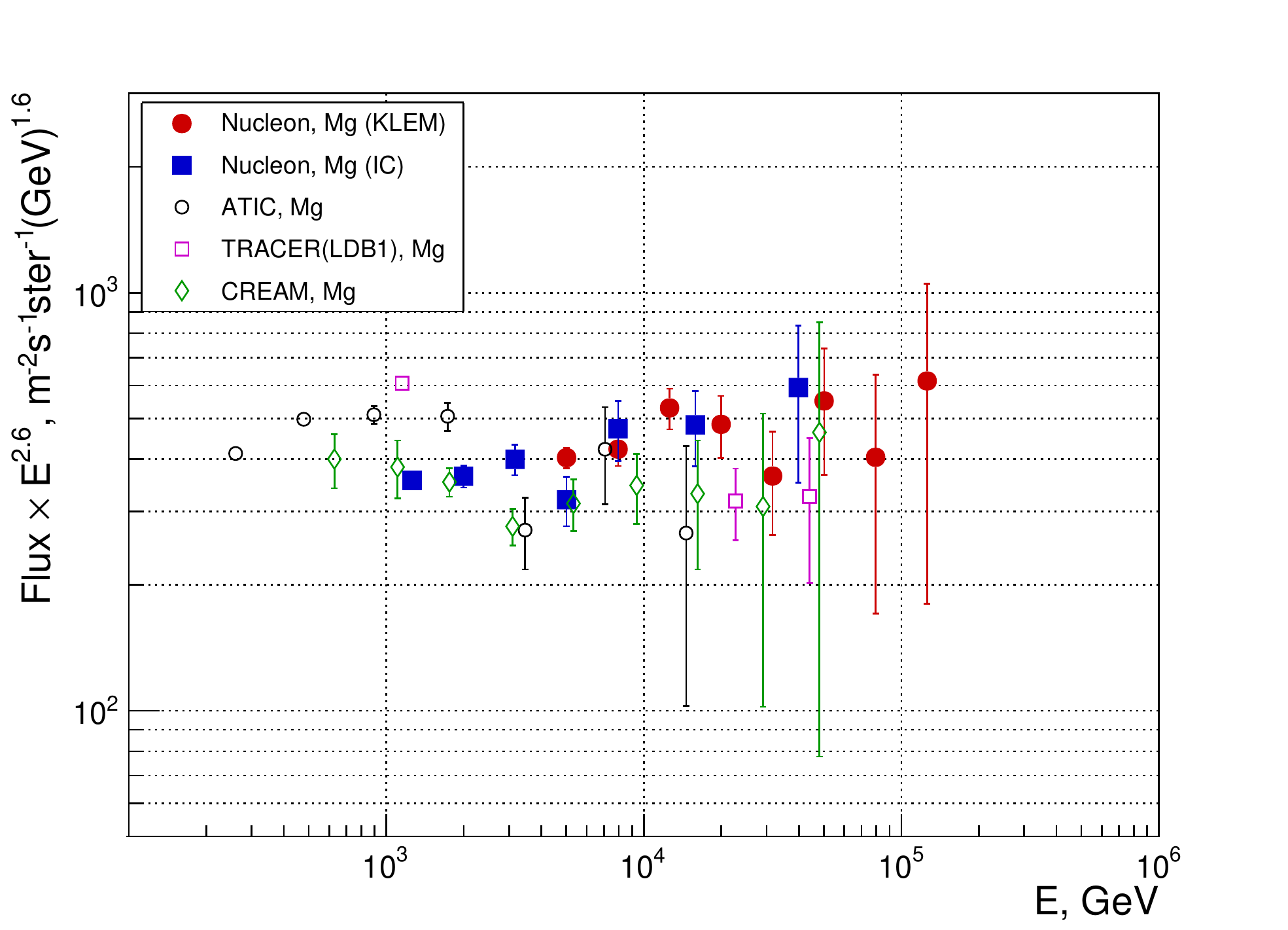}
\caption{\label{fig:NeMg} Energy spectra of neon and manganese obtained by the NUCLEON experiment and in the experiments: ATIC \cite{ATIC-2009-PANOV-IzvRAN-ENG}; TRACER(LDB1) \cite{CRNUCL-TRACER2008B-ApJ}; TRACER(LDB2) \cite{CRNUCL-TRACER2011-ApJ}; CREAM \cite{CR-CREAM2010A}.} 
\end{figure}

\begin{figure}
\centering
\includegraphics[width=0.75\textwidth]{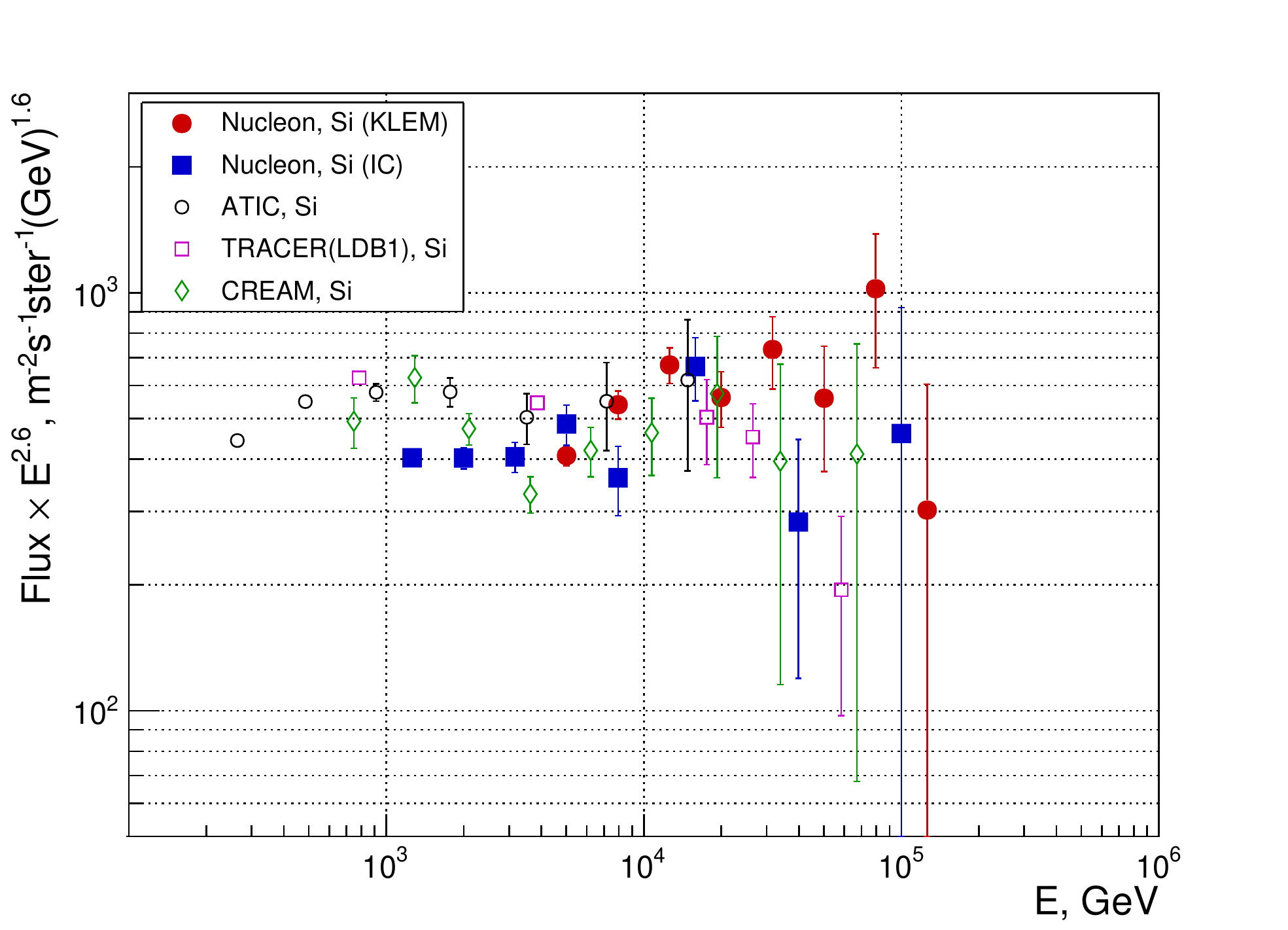}
\includegraphics[width=0.75\textwidth]{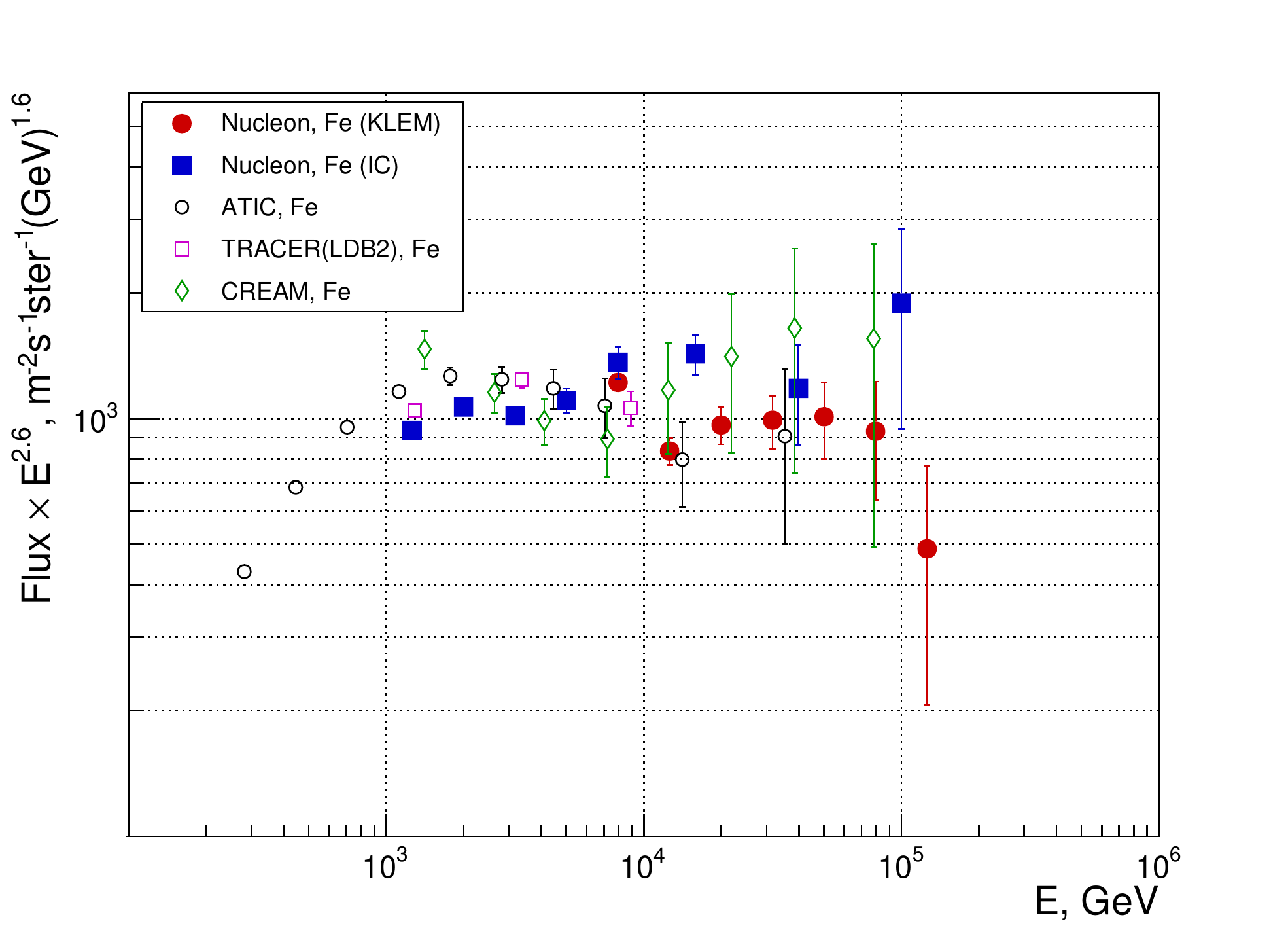}
\caption{\label{fig:SiFe} Energy spectra of silicon and iron obtained by the NUCLEON experiment and in the experiments: ATIC \cite{ATIC-2009-PANOV-IzvRAN-ENG}; TRACER(LDB1) \cite{CRNUCL-TRACER2008B-ApJ}; TRACER(LDB2) \cite{CRNUCL-TRACER2011-ApJ}; CREAM \cite{CR-CREAM2010A}.} 
\end{figure}

Figure~\ref{fig:CO} shows the energy spectra of carbon and oxygen nuclei obtained by the NUCLEON experiment; figures~\ref{fig:NeMg} and \ref{fig:SiFe} shows the energy spectra of neon, silicon, manganese and  iron nuclei. 
There are no strong deviations from the results of the other experiments (see the captions under the pictures). 
Some systematic differences between the calorimeter and the KLEM methods are present only for carbon and iron nuclei. 
For the heavy nuclei, the spectra obtained show several interesting features.

\begin{figure}
\centering
\includegraphics[width=0.75\textwidth]{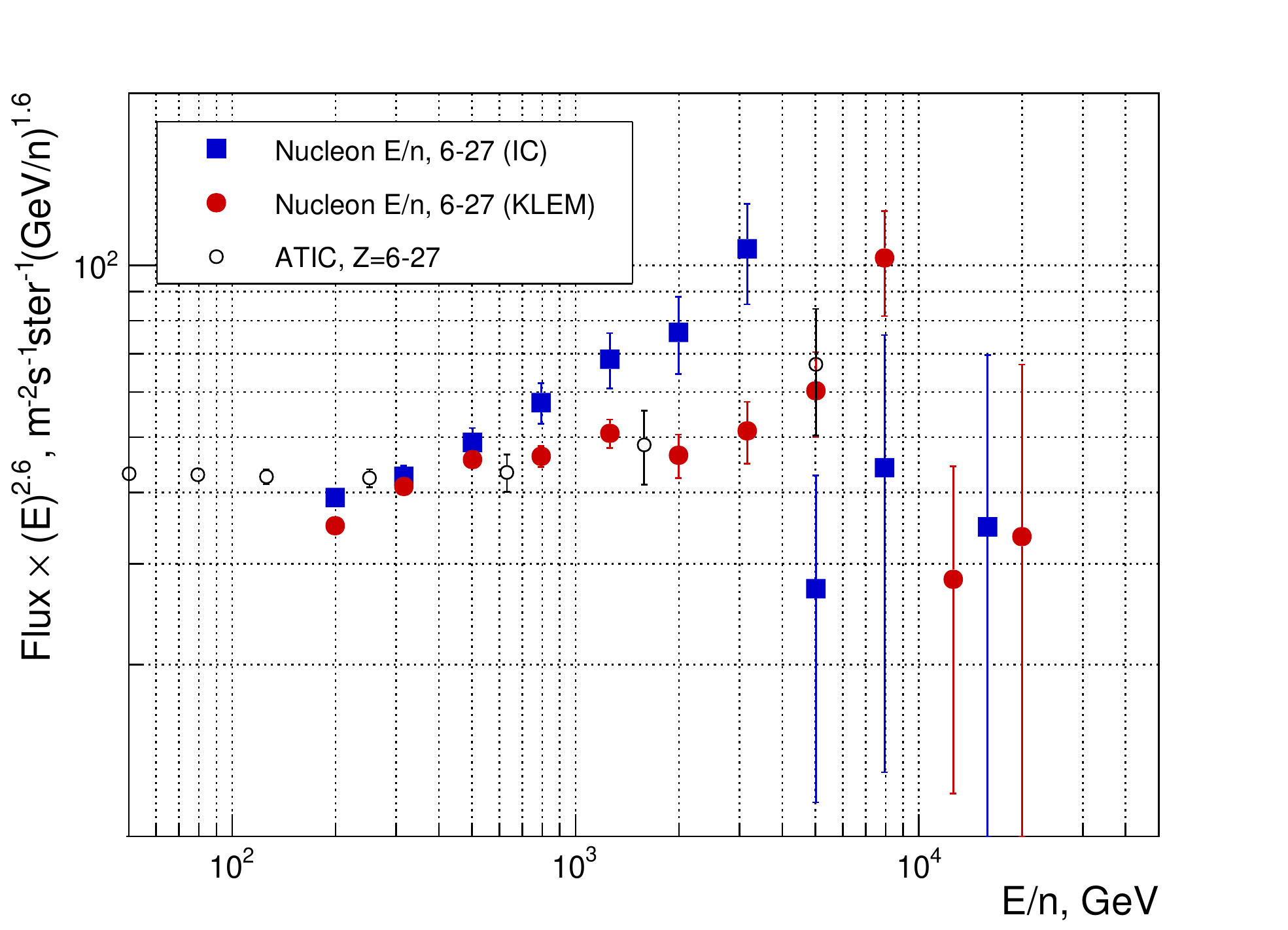}
\caption{\label{fig:AllNucl} Spectra of heavy nuclei ($Z=6\div27$) in terms of energy per nucleon from the NUCLEON experiment along with similar data from the ATIC experiment \cite{ATIC-2009-PANOV-IzvRAN-ENG}.} 
\end{figure}

One of the intriguing problems is the possibility of a flattening of the spectra for the majority of heavy nuclei at high energies -- above a few hundred GeV per nucleon. 
An indication of the existence of such a phenomenon was seen in the ATIC experiment \cite{ATIC-2007-PANOV-IzvRAN} and, later, in the CREAM experiment \cite{CR-CREAM2010A}. 
The TRACER experiment \cite{CRNUCL-TRACER2011-ApJ,CRNUCL-TRACER2008B-ApJ} did not confirm the existence of this effect, but it does not apparently contradict it due to insufficient statistical accuracy. 
Some indications of the existence of this phenomenon can be seen in the spectra from the carbon and oxygen nuclei (figure~\ref{fig:CO}), but it is absent from the iron spectrum (figure~\ref{fig:SiFe}). 
Significantly more reliable data can be obtained by constructing an averaged spectrum of the heavy nuclei in terms of energy per nucleon, which can dramatically increase the statistical significance of the spectrum at high energies. 
Figure~\ref{fig:AllNucl} shows the spectra of heavy nuclei ($Z=6\div27$) in terms of energy per nucleon from the NUCLEON experiment along with the similar data from the ATIC experiment \cite{ATIC-2009-PANOV-IzvRAN-ENG}. 
For historical reasons (for comparison with the data of the ATIC), the spectrum of the iron nucleus is also included, while the iron spectrum has no signs of flattening at high energies (as it will be specifically discussed below). 
Although there are some systematic differences in the absolute intensity of the spectrum between the calorimeter method and the KLEM method of the NUCLEON experiment, qualitatively, both methods reliably indicate that the averaged spectrum of heavy nuclei at energies above 200--300 GeV per nucleon has a low slope, confirming the indications of the ATIC experiment. 
However, the NUCLEON data also provide evidence of an entirely new phenomenon, which could not be detected in the ATIC experiment: at energies above 3--8\,TeV/n (depending on the method used) the spectrum unexpectedly goes down dramatically. 
This indication is not quite statistically robust, but it will be checked with the accumulated data of the NUCLEON experiment and improved data processing methods. 
Note that this phenomenon manifests itself in a previously inaccessible energy range, and the NUCLEON experiment was designed for the sake of such physics.

\subsection{Features of the spectrum of iron in comparison with the spectra of other heavy nuclei}
\label{subsec:Iron}

\begin{figure}
\centering
\includegraphics[width=0.75\textwidth]{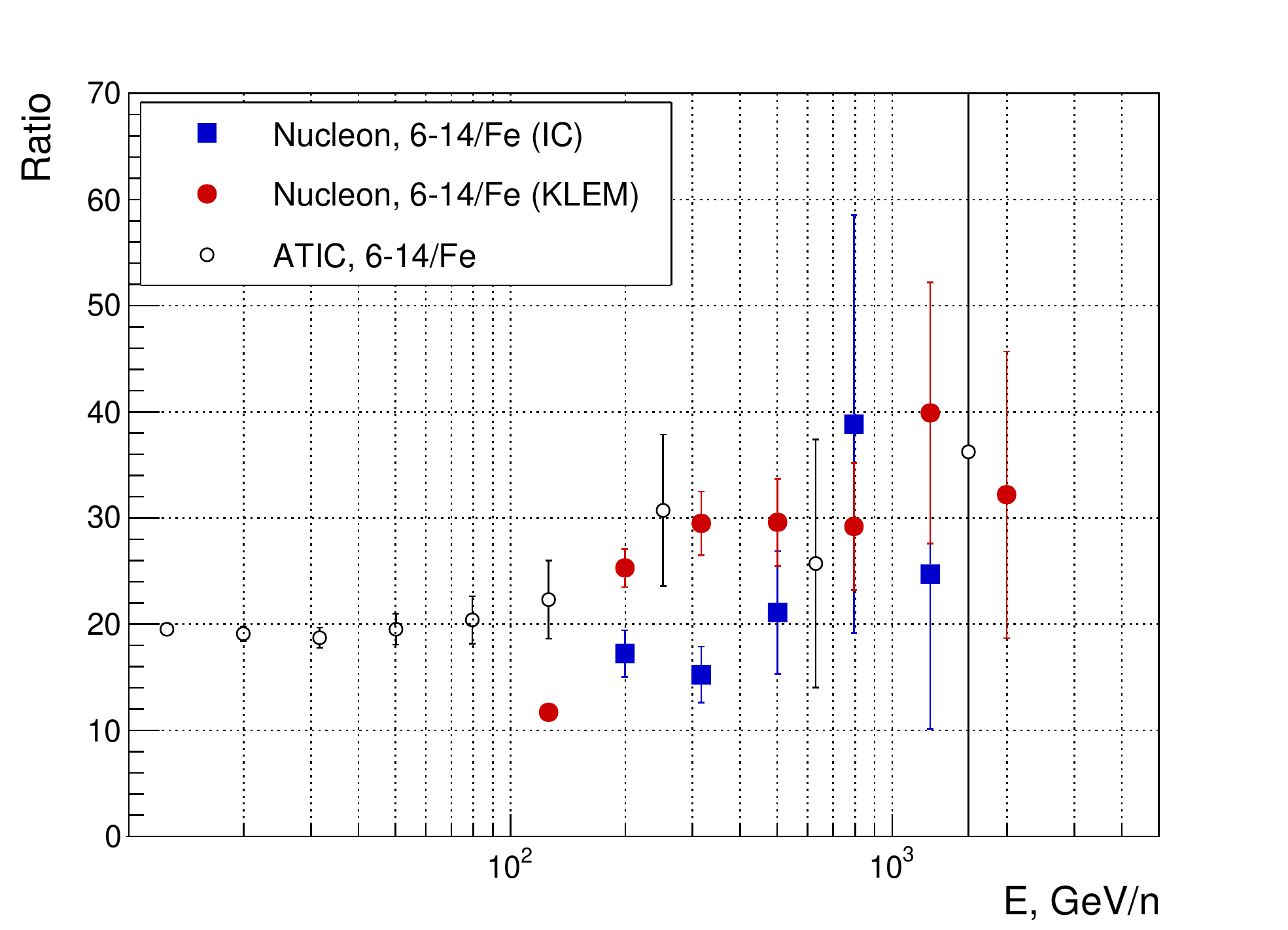}
\caption{\label{fig:Abund-to-Fe} Ratios of the spectra of nuclei with charges $Z=6\div14$ in terms of energy per nucleon to the spectrum of iron nuclei for the NUCLEON experiment and the ATIC experiment \cite{ATIC-2013-Panov-FeUpturn}.}
\end{figure}

As can already be seen based on the data presented, the iron spectrum behaves significantly differently from the spectra of other heavy nuclei at high energies. 
The easiest way to see this is to determine the ratios of the heavy nuclei spectra to the iron spectrum. 
That has already been done in the ATIC experiment \cite{ATIC-2014-NuclPhysB} and the results do indicate a significant difference between these spectra, although the statistical significance of the data is not very high. 
Those findings can be tested in the NUCLEON experiment with greater statistical reliability and for higher energies. 
Figure~\ref{fig:Abund-to-Fe} shows the ratios of the spectra of nuclei with charges from 6 to 14 in terms of the energy per nucleon to the spectrum of the iron nucleus for the NUCLEON experiment and the ATIC experiment \cite{ATIC-2014-NuclPhysB}. 
The NUCLEON data confidently indicate that the spectrum of iron at energies above $\sim$100\,GeV per nucleon is steeper than the spectra of heavy nuclei with charges from 6 to 14, which includes the abundant heavy nuclei C, O, Ne, Mg, Si. 
The systematic difference of the ratios obtained from the calorimeter method and the KLEM method is mainly caused by systematic differences in the measured spectrum of iron, already noted above. 
Qualitatively, however, both methods lead to the same result, and the statistical significance of the common result is mainly provided by the KLEM method, for which the statistics are about four times as many as the statistics of the calorimeter method. 
Note that the difference between the iron spectrum and the spectra of heavy nuclei currently has no explanation, which is why this phenomenon is very important. 
Also note that the difference between the spectra is observed in the NUCLEON experiment separately for each heavy nuclei and the iron nucleus, but the statistical significance of this difference is lower than for the total spectrum of 6--14 charges.

\begin{figure}
\centering
\includegraphics[width=0.75\textwidth]{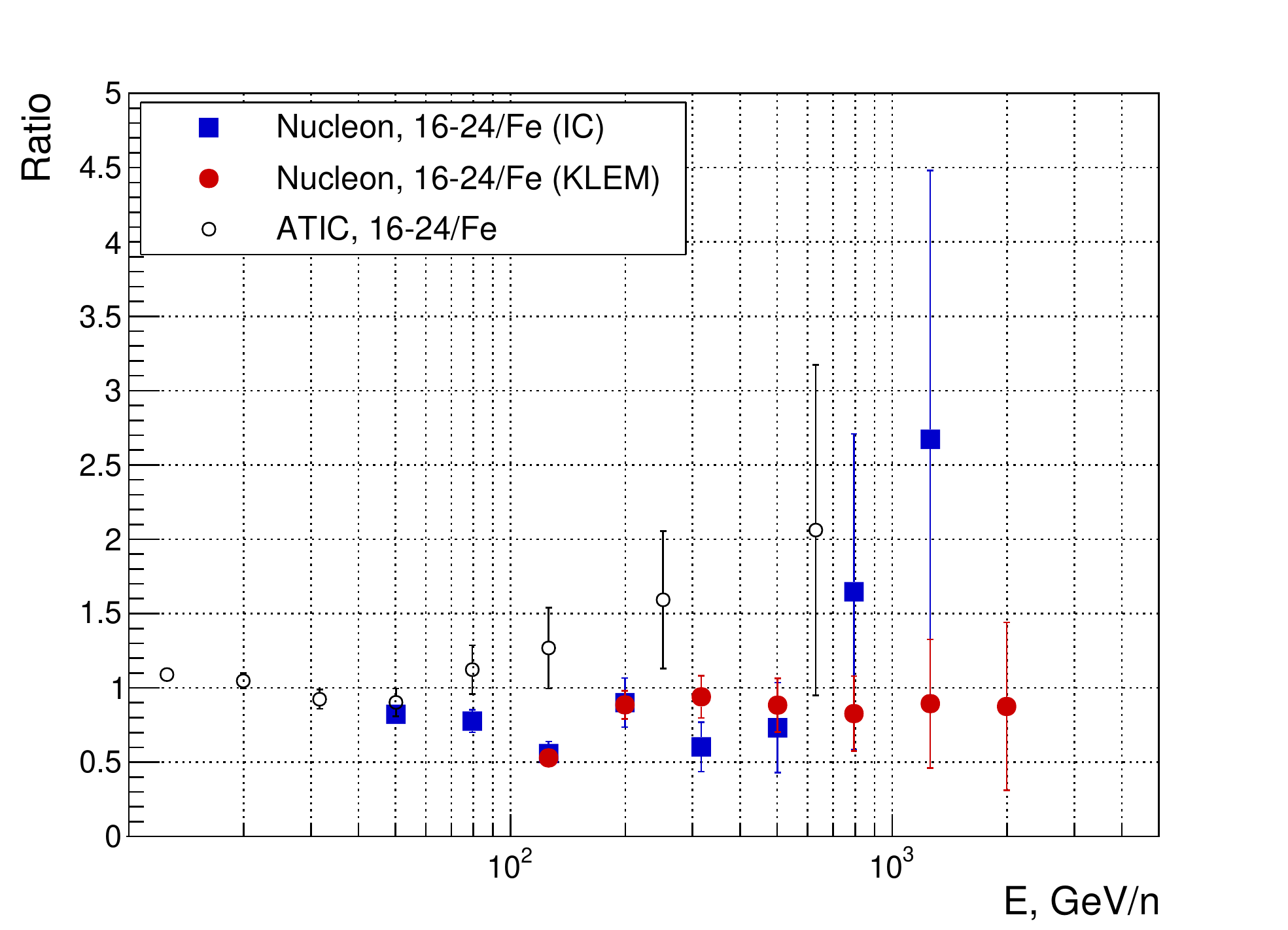}
\caption{\label{fig:SubFe-to-Fe} Ratios of the spectra of nuclei with charges $Z=16\div24$ (``sub-Fe'' nuclei) in terms of energy per nucleon to the spectrum of iron nuclei for the NUCLEON experiment and the ATIC experiment \cite{ATIC-2013-Panov-FeUpturn}.}
\end{figure}

The ATIC experiment indicates even greater differences between the spectra of heavy nuclei within the sub-Fe charge range $(Z = 16\div24)$ and iron nuclei (the iron spectrum is steeper). 
This is especially strange, since a lot of sub-Fe nuclei are secondary nuclei -- fragments from nuclear spallation, mainly of iron and interstellar gas, which are expected to have steeper spectra than the spectrum of iron. 
Earlier, a similar effect was observed in the results of the HEAO-3-C3 space experiment \cite{HEAO-HN-1987-ICRC-330}, but the authors then questioned the reality of the phenomenon and tied it to a possible methodological error. 
Figure~\ref{fig:SubFe-to-Fe} shows the ratios of the spectra of nuclei with charges from 16 to 24 (sub-Fe nuclei) in terms of energy per nucleon to the spectrum of iron nuclei for the NUCLEON experiment and the ATIC experiment \cite{ATIC-2013-Panov-FeUpturn,ATIC-2014-NuclPhysB}. 
As can be seen, there is a more mixed picture. 
The calorimeter method data qualitatively confirm the results from the ATIC experiment very well, but the statistical errors of the calorimeter method are large, as well as the statistical errors of the ATIC experiment. 
The KLEM method, although not explicitly showing the theoretically expected decrease of the $Z = 16\div24$\,/\,Fe ratio with energy, which is already important, does not show the growth of this ratio similar to the results and outcomes of the ATIC experiment and the calorimeter method of the NUCLEON experiment. 
It is difficult to talk about systematic differences between the results of the calorimeter and the KLEM methods of the NUCLEON experiment, because all the differences occur within the statistical uncertainty. 
The situation should become clearer with a larger set of data in the NUCLEON experiment. The methodological causes of such differences should also be carefully considered.

\section{Discussion and summary}
\label{sec:Discussion}

Although the NUCLEON experiment is in its initial phase, and the results so far are preliminary in nature, we can say with confidence that the data already give numerous indications of the existence of various non-canonical phenomena in the physics of cosmic rays, which are expressed in violation of a simple universal power law of the energy spectra. 
Some of the results confirm and essentially clarify the  data of earlier experiments. 
Worth mentioning here are: the difference between the slopes of the spectra of protons and helium; the difference between the spectra of heavy abundant nuclei and iron nuclei; the difference between the spectra of the sub-Fe nuclei $(Z = 16\div24)$ and that of iron nuclei; and the flattening of all the studied nuclei except the iron nuclei at energies above ~ 500 GeV/nucleon. 
These phenomena can be explained in terms of the concept of the existence of various sources of cosmic rays, which are characterized by different chemical compositions of the accelerated particles, and different energy spectra, such as in the three-component model \cite{CR-ZATSEPIN2006}, or within the concept of the heterogeneous structure of the cosmic ray sources themselves \cite{ATIC-2011-ZATSEPIN-ICRC,CRA-OHIRA2011}. 
The difference between the spectra of heavy nuclei and the spectrum of iron nuclei may be partly related to the effects of propagation in a heterogeneous space environment provided by the so-called ``superbubbles'' \cite{ATIC-2013-Panov-FeUpturn}, however, the discussion of these phenomena is in its initial phase. 
Other effects found (if confirmed) are brand new. 
These include breaks in the spectra of protons and helium near 10--20\,TeV per nucleon\footnote{During the revising of this paper, a message of the CREAM collaboration appeared \cite{CREAM2017-ApJ-pHe} that the CREAM experiment also confirms the existence of a break in the proton spectrum at energies of 10-20 TeV.} and a break in the spectra of heavy nuclei near the energies of 5--10\,TeV per nucleon. 
The existence of these phenomena is still not firmly established, so a discussion of their physical nature has not even begun. 
These phenomena are situated in a poorly investigated energy range, from 10\,TeV per particle energies up to several hundred TeV per particle, which became available in the NUCLEON experiment. 
They manifest themselves with the current amount of collected data and it is expected that the statistical significance of the results and their methodological elaboration will increase significantly during the experiment.

\acknowledgments

We are grateful to ROSCOSMOS State Space Corporation and Russian Academy of Sciences for their continued support of this research.

\end{document}